\documentclass{aa}
\usepackage{graphicx}
\usepackage{amsmath}
\usepackage{txfonts}
\usepackage{lscape}
\usepackage{rotating}
\usepackage{appendix}
\usepackage{url}
\usepackage{float}  
\usepackage{longtable, lscape}
\newcommand{\msun}{\mbox{\rm $M_{\odot}$}}
\newcommand{\msuns}{\mbox{\rm $M_{\odot}$}~}

\newcommand{\lsun}{\mbox{\rm $L_{\odot}$}}
\newcommand{\lsuns}{\mbox{\rm $L_{\odot}$}~}

\newcommand{\arcs}{\hbox{$^{\prime\prime}$}}

\newcommand{\arcm}{\mbox{$^{\prime}$}}

\newcommand{\into}{\mbox{$\times$~}}

\newcommand{\jhk}{\mbox{$JHK_{\rm s}$}~}
\newcommand{\ks}{\mbox{$K_{\rm s}$~}}

\newcommand{\degree}{\mbox{$^{\circ}$}}
\newcommand{\av}{\mbox{$A_{\rm V}$~}}

\newcommand{\rv}{\mbox{$R_{\rm V}$~}}
\newcommand{\hii}{\mbox{H~{\sc ii}~}}

\newcommand{\mum}{\hbox{$\mu$m}~}
\newcommand{\wat}{\hbox{H$_2$O}~}

\newcommand{\mjpb}{\hbox{~mJy~beam$^{-1}$}}

\newcommand{\tco}{\hbox{$^{13}$CO}~}

\newcommand{\hco}{\mbox{HCO$^+$}~}

\newcommand{\vlsr}{\mbox{$V_{\mbox{\tiny LSR}}$~}}

\def\HII{H\,{\sc{ii}}}

\def\h{\hbox{$^{\reset@font\r@mn{h}}$}}
\def\m{\hbox{$^{\reset@font\r@mn{m}}$}}
\def\s{\hbox{$^{\reset@font\r@mn{s}}$}}
\def\msol{\hbox{\kern 0.20em $M_\odot$}}

\def\kms{\hbox{\kern 0.20em km\kern 0.20em s$^{-1}$}}
\def\cmmt{\hbox{\kern 0.20em cm$^{-3}$}}
\def\cmsq{\hbox{\kern 0.20em cm$^{-2}$}}
\def\pc{\hbox{\kern 0.20em pc$^{2}$}}

\def\h13cop{\hbox{H$^{13}$CO$^{+}$}}

\begin{document}
   \title{The molecular complex associated with the  Galactic \hii region Sh2-90: a possible site of triggered star formation.}
%
\author{M. R. Samal \inst{1}
         \and A. Zavagno\inst{1}
         \and L. Deharveng\inst{1}
         \and S. Molinari\inst{2}
         \and D. K. Ojha\inst{3}
         \and D. Paradis\inst{4,5}
         \and J. Tig\'e\inst{1}
         \and A. K. Pandey\inst{6}
         \and D. Russeil\inst{1} }

\offprints{manash.samal@oamp.fr }
   \institute{Aix Marseille Universit\'e, CNRS, LAM (Laboratoire d'Astrophysique de Marseille) UMR 7326, 13388 Marseille, France 
         \and INAF – Instituto Fisica Spazio Interplanetario, via Fosso del Cavaliere 100, 00133 Roma, Italy 
         \and Department of Astronomy and Astrophysics, Tata Institute of Fundamental Research, Homi Bhabha Road, Mumbai 400 005, India 
         \and Universit\'e de Toulouse, UPS-OMP, IRAP, Toulouse, France 
          \and CNRS; IRAP; 9 Av. du Colonel Roche, BP 44346, F-31028, Toulouse, cedex 4, France
         \and Aryabhatta Research Institute of Observational Sciences, Nainital 263 129, India
    }             
   \date{Received ; Accepted }

  \abstract
    {}
   {We investigate the star formation activity in the molecular complex associated with the Galactic \hii region Sh2-90.}
   {We obtain the distribution of the ionized gas using radio-continuum maps obtained at 1280 MHz and 610 MHz, and the distribution of
   dense neutral material using  Herschel Hi-GAL observations.
   We use deep near-infrared and  Spitzer data to investigate the stellar content of the complex and identify the young stellar objects (YSOs).
   We discuss the evolutionary status of embedded massive YSOs (MYSOs) using their spectral energy distribution.}
   {The Sh2-90 region presents a bubble morphology  in the mid-infrared (size $\sim$  0.9 pc
  $\times$ 1.6 pc). Radio observations suggest it is an evolved \hii region with an electron density $\sim$ 144 cm$^{-3}$, emission measure $\sim$ 6.7 $\times$ 10$^{4}$ cm$^{-6}$ pc 
  and an ionized mass $\sim$ 55 \msun. From Hi-GAL observations it has been found that the \hii region is part of an elongated extended  molecular cloud (size $\sim$ 5.6 pc $\times$ 9.7 pc, ${\rm H}_2$ column density $\geq$ 3 $\times$ 10$^{21}$ cm$^{-2}$, and  dust temperature 18-27 K) of total mass $\geq$ 1 $\times$ 10$^4$ \msun. We identify the ionizing cluster of Sh2-90, the main exciting star being an O8--O9 V star. 
  Five cold dust clumps (mass $\sim$ 8-95 \msun), four mid-IR blobs around B stars, and a compact \hii region are found at the edge of the bubble.
  The velocity information  derived from CO (J=3-2) data cubes suggests that most of them are associated with the Sh2-90 region.
  One hundred and twenty-nine YSOs have been identified (Class I, Class II, and near-IR excess sources). 
The majority of the YSOs are low-mass ($\leq$ 3 \msun) sources and they are distributed mostly  in the regions 
  of high column density. Four candidate  Class 0/I MYSOs have been found; they will possibly evolve to stars of mass $\geq$ 15 \msun. 
We suggest multi-generation star formation is present  in the complex. 
  From the evidence of interaction, the time scales involved, and the evolutionary status of stellar/protostellar sources, we argue that the star formation at the immediate border/edges of Sh2-90 
  might have been triggered by the expanding \hii region. However, several young sources in this complex are probably formed by some other processes.}
 {}

\titlerunning{Triggered star formation at the border of Sh2-90}
\authorrunning{M. R. Samal et al.}
\maketitle


\section{Introduction\label{intro} }
There are several ways 
the energy inputs from the OB  stars can modify the physical environment and  chemistry of the host complex in which 
they reside  (e.g., McKee \& Ostriker 1977), and therefore can trigger the formation of a new generation of stars in the 
complex  (e.g., Elmegreen \& Lada 1977; Bertoldi \& McKee 1990). 
 Recent observational studies of bubbles associated with \hii regions (e.g., Deharveng et al. 2010), suggest that their expansion possibly 
 triggers 14\% to 30\% of the star formation in our Galaxy (e.g., Deharveng et al. 2010; Thompson et al. 2012; Kendrew et al. 2012). These observational results have revealed  
the importance of OB stars on star formation activity on a Galactic scale. The detailed studies of individual objects 
(e.g., Deharveng et al. 2006, 2008; Urquhart et al. 2006; Zavagno et al. 2006, 2007), however, showed that the complex environments around \hii regions makes determining the exact process of star formation difficult and that, in general, this process is complicated. 

  Now the far-infrared (FIR) observations 
provided by Hi-GAL ({\it Herschel} Infrared Galactic Plane Survey; Molinari et al. 2010a) using the {\it Herschel} Space Observatory
have the ability to image a large area of a cloud complex 
with unprecedented sensitivity, thus allowing us to improve our understanding of how OB stars interact 
with the local interstellar medium (ISM), and process cold gas to induce new star formation. 
The recent studies based on  {\it Herschel} observations revealed that 
star-forming regions (SFRs) are composed of very complex and filamentary  clouds, 
with ongoing star formation at various locations (e.g., Hill et al. 2011; Giannini et al. 2012;  Hennemann et al. 2012, 
Deharveng et al. 2012; Schneider et al. 2012; Preibisch et al. 2013;  Roccatagliata et al. 2013).
Demonstrating the existence of triggered star formation in extended clumpy clouds by  internal feedback sources 
is difficult, because  they may be forming new stars in various ways. 
For example, such clouds can form stars spontaneously governed by the physical condition and the evolution 
of the original cloud   
or the  collapse of high-density structures generated by the large-scale supersonic turbulence of the the ISM (e.g., see Mac Low \& Klessen 2004). Thus, understanding of 
the physical connection and interaction of bubbles/\hii regions with the cold  ISM, 
and their association with stellar/proto-stellar content  is central to obtaining a better picture of star formation around \hii regions. 
In this context, we present here a multiwavelength investigation of the Sh2-90 \hii complex (briefly described in Sect. 2) 
in order to decipher its  star formation activity.

In the present work,  we analyzed the distributions of the ionized and cold neutral ISM,
with radio continuum observations at low frequencies (610 and 1280 MHz)
and dust continuum emission with {\it Herschel} in the wavelength range 70-500 $\mu$m.
We explore the  stellar and proto-stellar components of the complex, using
high-resolution $JHK$ observations coupled with archival {\it Spitzer}-IRAC observations.
We have organized this work as follows. Section 2 presents the Galactic \hii region Sh2-90. In Sect. 3, we describe the observations and the 
reduction procedures. Section 4 describes the \hii region (adopted distance, general morphology, properties of ionizing gas, and exciting source).  
Section 5 deals with the distribution and physical condition of the cold  ISM, and the properties of compact dusty clumps.  In Sect. 6,
we describe the identification and classification of  young stellar objects (YSOs), their nature and distribution.
The kinematics of ionized and molecular gas is presented in Sect. 7. Section 8 is devoted to the general discussion and
our understanding of star formation in the Sh2-90 complex. We present the main conclusions in Sect. 9.

\begin{table*}
\centering
\scriptsize
\caption{ IRAS point sources towards the Sh2-90 complex}
\begin{tabular}{cccccccc}
 \hline\hline
 Name  & RA  & Dec & F$_{12}$ & F$_{25}$ & F$_{60}$ & F$_{100}$ & L  \\
  & deg (J2000)   & deg (J2000) & Jy & Jy & Jy & Jy & 10$^3$  \lsuns  \\
  \hline
  IRAS 19473+2638 & 297.341461 & 26.775908 & 6.02 & 14.25 & 111.90 & 1514.00 & 5.7 \\
  IRAS 19474+2637 & 297.385773 & 26.753876 & 4.20 & 14.75 & 92.02 & 3193.00 & 10.4\\
  IRAS 19473+2647 & 297.358250 & 26.920030 & 2.62 & 1.40 & 3.31 & 49.50 & 0.3\\
  IRAS 19471+2641 & 297.291168 & 26.814301 & 7.68 & 29.29 & 285.80 & 1673.00 & 7.7\\
  IRAS 19470+2643 & 297.287231 & 26.849007 & 2.93  & 44.75 &  330.60 & 1514.00 & 7.6 \\
 \hline
\end{tabular}
\end{table*}
\begin{figure*}
\center
  \includegraphics[width=13cm,height=13cm ]{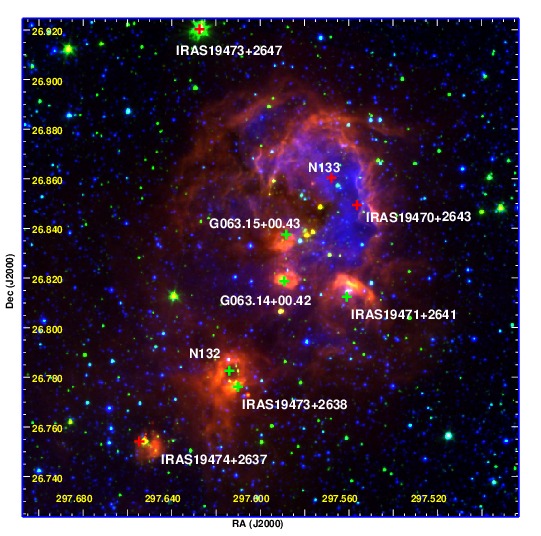}
 \caption{Colour-composite image of the Sh2-90 complex. {\it Spitzer}-IRAC 8.0 \mum
  (red) and 3.6 \mum (green) images have  been combined with the 
  DSS2 R-band (blue) image. The different sources 
  associated with the region (see the text) are marked. The field
  size is 12$\farcm$0 (E-W) $\times$ 12$\farcm$0 (N-S), centered at
 $\alpha_{2000} = 19^{\rm h}49^{\rm m}18^{\rm s}$,
$\delta_{2000} = +26^{\circ}49^{\prime}29^{\prime\prime}$. North is
  up and east is  left.}
  \label{fig1}
\end{figure*}

\section{Description of the Sh2-90 complex}
The Sh2-90 complex (Sharpless 1959), located at $\alpha_{2000} = 19^{\rm h}49^{\rm m}11^{\rm s}$, $\delta_{2000} =  +26^{\circ}51^{\prime}36^{\prime\prime}$ 
 ($l = 63\degree.16$, $b = 0\degree.40$),  is an optically visible irregularly shaped \hii region.
This \hii region is a part of the Vulpecula OB association (Turner 1986). The  most commonly adopted distances
to the \hii region  are between 1.6 kpc and 2.5 kpc (Strak 1984; Beaumont \& Williams 2010; Russeil et al. 2011).  We
discuss the distance in Section 4 and adopt a value of 2.3 kpc in the present study. The main exciting source of the nebula 
is uncertain; measurements based on an indirect approach suggest that it has been created by a star of O9.5V-O8V spectral type (Georgelin 1975; Lafon et al. 1983).
Observations in the \tco (110.201 GHz) and \hco (89.189 GHz) lines suggest that 
Sh2-90 is a part of a massive ($\sim$ 4 $\times$ 10$^4$ \msun) asymmetric cloud
(Lafon et al. 1983). 

The Sh2-90 molecular cloud complex contains several kinds of sources. Figure 1 displays the colour-composite image made with  the R-band (DSS2 survey) in blue, the
emission at 3.6~$\mu$m in green and the  8.0~$\mu$m emission
in red ({\it Spitzer}-GLIMPSE survey; see Sect. 3.3). The various sources discussed 
in the present work are marked  in  
Fig. 1. The sources N133 and N132  are identified as bubbles in the {\it Spitzer} 8.0 \mum band  (Churchwell et al. 2006) of the GLIMPSE survey. 
Of the two bubbles, N133 is an elliptical bubble of average radius $\sim$ 1$\farcm$6, which encloses the \hii region Sh2-90, while N132 is a circular bubble of average radius $\sim$ 0$\farcm$28; N133 is associated with a detectable radio \hii region (Israel 1977), but no radio emission has been reported in the direction 
of N132. Several IRAS sources (see Table 1) are identified in  close proximity to Sh2-90. 
The fluxes of these IRAS sources at 25, 60, and 100 \mum have been taken from the IRAS Catalog of Point Sources, Version 2.0 (Helou \& Walker 1988). We estimated the FIR luminosity (see Table 1) of these IRAS sources  from the IRAS fluxes  using the relation given 
in Casoli et al. (1986). The source IRAS 19473+2638 (luminosity $\sim$ 5.7 $\times$ 10$^3$ \lsun) is located in 
the close vicinity of N132, whereas IRAS 19474+2637 (luminosity $\sim$ 1.0 $\times$ 10$^{4}$ \lsun) is located  
$\sim$ 2$\farcm$3 S-E of N132. 
The location of IRAS 19473+2647 is about $\sim$ 5$\farcm$0 N-E of N133, and is a low luminosity ($\sim$ 0.3 $\times$ 10$^3$ \lsun)
source.

Apart from the above distinct IRAS sources, close to Sh2-90 and at its southern edge lies the 
IRAS source 19474+2641 of luminosity $\sim$ 7.7 $\times$ 10$^3$ \lsun. 
Another IRAS source 19470+2643  of  luminosity $\sim$ 7.6 $\times$ 10$^3$ \lsuns 
is  situated at the western edge of Sh2-90. 
Figure 1 also shows two compact 
symmetric 8.0 $\mu$m dust emissions at the eastern side of  Sh2-90. 
These symmetric 8.0 $\mu$m structures coincide with the MSX point sources (Egan et al. 2003)
G063.1549+00.4309 and G063.1422+00.4227. The colours ($F_{\rm 21}$/$F_{\rm 8}$ and $F_{\rm 14}$/$F_{\rm 12}$) of 
these sources at MSX bands (8.28 $\mu$m (A), 12.13 $\mu$m (C), 14.65 $\mu$m (D), and 21.34 $\mu$m (E)) fall in the
criteria of a compact \hii region developed by Lumsden et al. (2002), who used these colours to identify 
various sources in the Galactic Plane. 
Altogether, the Sh2-90 region has a bubble morphology with 
minimal 8 $\mu$m emission at its center, and is surrounded by several luminous Infrared (IR) sources. These 
configurations are possible sites of a new generation induce star formation (e.g., Deharveng et al. 2005; Zavagno et al. 2007; Samal et al. 2012). 
Thus, the Sh2-90 complex is a potential target for examining the influence of an \hii region on  star formation processes.

\section{Observations and data reduction}
\subsection{Optical photometry}
The optical photometric  observations at the $V$ and $I$ bands  were  performed  for  the
Sh2-90 region (centered on $\alpha_{2000} = 19^{\rm h}49^{\rm m}29^{\rm s}$, $\delta_{2000} =  +26^{\circ}50^{\prime}13^{\prime\prime}$) on 2006 June 02, using the 2K
$\times$ 2K  CCD system of the 104 cm Sampurnanand telescope, 
Nainital (India).   The 0.37
arcsec pixel$^{-1}$ plate scale  gives a field of view (FoV) of
$\sim$ $12\farcm6 \times 12\farcm6$ on  the sky.
To improve the signal-to-noise ratio (S/N), the
observations were carried out in binning mode of 2 $\times$ 2 pixels.
The observing conditions were photometric and the average FWHM
during the observing period was $\sim$ 1$\farcs$7-
2$\farcs$0. The initial processing and photometry of the
images were done using  IRAF. We used the point spread function (PSF) algorithm 
ALLSTAR in the DAOPHOT package to extract
the photometric magnitudes.
The PSF was determined from the bright and isolated stars of the field.
 The calibration from instrumental 
to standard system was done using the procedure outlined by 
Stetson (1987), using the standard field SA101  (Landolt 1992) 
observed during the same night. The standardization residuals between the
standard and transformed $V$ and $I$ magnitudes and colours were less than 
0.05 mag. Stars having photometric error  $\leq 0.1 $ mag
are used in the present analyses.

Completeness limits of the observations were calculated  quantitatively by  plotting 
histograms of the point sources, and we considered that the data is complete up to 
the linear distribution in the histograms. With this approach  the approximate completeness 
limits for the V- and I- band data are 18.0 mag and 18.5 mag, respectively.

\subsection{CFHT near-infrared imaging}
 Deep NIR observations of the Sh2-90 region
(centered on $\alpha_{2000} = 19^{\rm h}49^{\rm m}18^{\rm s}$,
$\delta_{2000} = +26^{\circ}48^{\prime}46^{\prime\prime}$)
 in the $J$ ($\lambda$ = 1.25  $\mu$m), $H$ ($\lambda$ = 1.63
$\mu$m), and  \ks ($\lambda$ = 2.14  $\mu$m) bands were obtained on 2006 
July 06 with the  WIRCAM camera of the CFHT 3.6 m telescope (Puget et al. 2004).
In this set-up,
each pixel corresponds to 0$\farcs$3 and
yields a FoV $\sim$ 20$\farcm$0 $\times$ 20$\farcm$0  on the sky.
The observing conditions were photometric and
the average FWHM during the observing period was $\sim$0$\farcs$7-0$\farcs$9. The initial processing of 
the data was done in the CFHT pipeline software TERAPIX (Bertin et al., 2002). 
We perform photometry for an area of $\sim$ 12$\farcm$5 $\times$ 12$\farcm$5 centered on
 $\alpha_{2000} = 19^{\rm h}49^{\rm m}18^{\rm s}$,
$\delta_{2000} = +26^{\circ}49^{\prime}29^{\prime\prime}$. 
Photometry on the images  was done using the  PSF algorithm of DAOPHOT package (Stetson 1987) in   IRAF.
The PSF was determined from the bright and isolated stars of the field.
For photometric calibration, we used isolated Two Micron
All Sky Survey (2MASS) point sources (Cutri et al. 2003) having error $<$ 0.1 mag and rd-flag ``123".
Rd-falg values of 1, 2 or 3 generally indicate the best quality detections, photometry, and astrometry. 
A mean calibration dispersion $\leq$ 0.06 mag is observed in each band, 
indicating that our photometry is reliable within ∼0.06 mag; Two hundred twenty-nine sources were 
found to be saturated in our catalog; these sources were replaced by 2MASS sources. 
Sources with photometric error  $\leq 0.1 $  mag 
in all the three bands  are considered in the present work.

To evaluate the completeness of the census of JHK detection quantitatively, we plot histograms of the JHK point sources and  
considered the data is complete up to the linear distribution in the histograms (e.g., Ohlendorf et al. 2013). In this way the completeness 
limits for the J-, H-, and K-band data are $\sim$ 19, 18, and 17 mag, respectively.

\subsection{Spitzer observations and point source catalogs}
 
The  Sh2-90 complex was observed as part of the Galactic Legacy Infrared Mid-Plane Survey Extraordinaire (GLIMPSE; Benjamin et al. 2003; PI: E. Churchwell; Program ID: 188)  and 
the Multiband Imaging Photometer GALactic plane survey (MIPSGAL; Carey et al. 2009; PI: S. Carey; Program ID: 20597) by {\it Spitzer} Space Telescope. 
We downloaded the Post Basic Calibrated Data (PBCD) images of the {\it Spitzer} Infrared Array Camera (IRAC) at 3.6, 4.5, 5.8, and 8.0 $\mu$m,  and Multiband Imaging Photometer (MIPS)
24.0 $\mu$m PBCD images from the {\it Spitzer} Archive\footnote{http://sha.ipac.caltech.edu/applications/Spitzer/SHA/} to study the morphology of the complex. 
The angular resolution of the  images at IRAC bands  are $<$ 2\farcs0, whereas it is  $\sim$ 6\farcs0 at the MIPS 24 $\mu$m band.

For the point source analyses, we used the GLIMPSE point source catalog  available on the 
Vizier web site\footnote{http://vizier.u-strasbg.fr/viz-bin/VizieR?-source=II/293}. Applying the same procedure as described in Section 3.2, we quantitatively considered that the 
IRAC point source catalog is nearly complete down to 13.5, 13.5, 12.0, and 11.0 mag at 3.6, 4.5, 5.8, and 8.0 $\mu$m bands, 
respectively.

\subsection{{\it Herschel} multi-band observations in the range  70-500 \mum}
The far-infrared data used in this paper were taken with the {\it Herschel}-PACS (Poglitsch et al. 2010) 
and SPIRE (Griffin et al. 2010) imaging cameras as part of
the Hi-GAL survey. 
This  survey is a {\it Herschel} Open Time key
project that  maps the whole Galactic Plane in five bands centered at 70 $\mu$m and 160 $\mu$m with
PACS, and 250 $\mu$m, 350 $\mu$m, and 500 $\mu$m with SPIRE (Molinari et al. 2010a, b). 
The spatial resolutions of these bands are 6$\farcs$7, 11$\farcs$0, 18$\farcs$0, 25$\farcs$0, and 37$\farcs$0,
respectively.  At the distance of the Sh2-90 complex, this corresponds to a physical scale in the range  0.07-0.41 pc.
The data are acquired in PACS/SPIRE Parallel mode by moving the satellite at
a constant speed of 60$\farcs$0/sec and acquiring images simultaneously
in the five photometric bands. The detailed description of the observation
settings and scanning strategy adopted is given in
Molinari et al. (2010b). The detailed description
of the pre-processing of the data up to usable  high-quality maps can be found in 
Traficante et al. (2011). 

\subsection{Radio continuum mapping}

The radio continuum interferometric observations of the Sh2-90 region (centered on $\alpha_{2000} = 19^{\rm h}49^{\rm m}18^{\rm s}$,
$\delta_{2000} = +26^{\circ}53^{\prime}12^{\prime\prime}$) at 1280 MHz and 610 MHz were carried out on 2007 December 22 
and 2007 December 28, respectively, using the Giant Metrewave Radio Telescope (GMRT) array (Swarup et al. 1991).  
The sources 3C48 and 3C286,  and 1924+334 were observed 
as flux and phase calibrators to derive the phase and amplitude gains.
The NRAO Astronomical Image Processing System (AIPS) was used for the data reduction.
While calibrating the data, the data were carefully checked and bad data points were flagged at various stages.
The estimated uncertainty of the flux calibration was within
8\% at both frequencies. The image of the field was formed  by Fourier inversion and
the cleaning algorithm task IMAGR. The resulting images at 1280 MHz (beam $\sim$ 16$^{\prime\prime}$ \into
11$^{\prime\prime}$, rms noise $\sim$ 1.2 mJy/beam) and at 610 MHz (beam $\sim$ 34$^{\prime\prime}$ \into 22$^{\prime\prime}$, 
rms noise $\sim$ 7 mJy/beam)  were  made with a Brigg's weighting function (robust factor 0). 
At the distance of the Sh2-90 complex, the mean beam width of the radio images corresponds 
to a physical scale in the range 0.15- 0.31 pc.
Few iterations of  self-calibration were 
carried out to remove the residual effects of atmospheric and ionospheric 
phase corruptions and to obtain improved maps. The system temperature 
correction was done using sky temperature from the 408 MHz map of Haslam et
al. (1990). A correction factor equal to the ratio of the system
temperature toward the source and flux calibrator has been used to
scale the images.

For the Sh2-90 complex, different observations with different areas have been conducted or taken from the archive, but in the following 
we have done the point source analyses   for  a common  area of $\sim$ 12\arcm $\times$ 12\arcm centered on
 $\alpha_{2000} = 19^{\rm h}49^{\rm m}18^{\rm s}$,
$\delta_{2000} = +26^{\circ}49^{\prime}29^{\prime\prime}$ (i.e., the area of Fig. 1).

\begin{table}
\caption{Kinematic information of the Sh2-90 complex}
\begin{tabular}{ccc}
 \hline\hline
  line   & \vlsr (\kms)  & references  \\
  \hline

  CO       & 22.2           & Blitz et al.\ (1982) \\
 $^{13}$CO (1-0)     & 20.5 & Lafon  et al. (1983) \\
  CO  (2-1)      & 21.0      & Lafon  et al. (1983) \\
 $^{18}$CO (1-0)  & 20.5      & Lafon  et al. (1983) \\
 HCO$^+$ (1-0)      & 22.0   & Lafon  et al. (1983) \\
 CO(3-2)       & 21.1           & Beaumont \& Williams (2010) \\

 \hline
\end{tabular}
\end{table}

\section{Distance, morphology, and nature of the ionized gas}
The distance of the region is uncertain and different indicators suggest  a near kinematic distance in the range of 1.6-2.5 kpc 
(for discussion, see Lafon et al. 1983; Beaumont \& Williams 2010; Russeil et al. 2011). 

The mean velocities of the molecular  gas observed towards the Sh2-90 \hii region ($l$=63\degree.10 $b$=+0\degree.46) are given in Table 2. Table 2 shows that the 
molecular emission towards the region is mainly  in the velocity range 20.5-22.2 \kms. Using the Galactic rotation curve of Brand \& Blitz (1993), this velocity range 
corresponds to a near kinematic distance in the range  2.1-2.4 kpc.

Lafon et al. (1983), based on the [O{\sc III}/H$_\beta$] strength, suggested that the main exciting source of Sh2-90 is a 
star of spectral type O8-O9. We identify the exciting source 
at the center of the nebula (see Sect. 4.3). Using its near-infrared magnitude and colour (J = 11.00, J-H=0.99 and J-K =1.46), 
the synthetic photometry of O stars by Martins \& Plez (2006), and the interstellar extinction law of
Rieke \& Lebofsky (1985), we estimated a distance of 2.5 kpc for an O8V star 
and 2.2 kpc for an O9V star. 

From the above analyses it appears that the distance of Sh2-90  lies in the
range of 2.1-2.5 kpc. In the following, we thus adopted a mean distance of 2.3 $\pm$ 0.2 kpc.
From our radio observations, with this distance, we  derived the Lyman continuum photons emanating from
the associated massive star of the region (see  Sect. 4.2), which is compatible with an  O8-O9 star. 
This further supports the distance of 2.3 kpc for the region. We note that at this distance 
an angular size 1\farcs0 corresponds to a physical size $\simeq$ 0.01 pc on the sky.

\subsection{Infrared and radio  view of the complex}
\begin{figure}[htp]
\center
  \includegraphics[trim=0cm 0cm 0cm 0cm,angle=0,width=8.0 cm ]{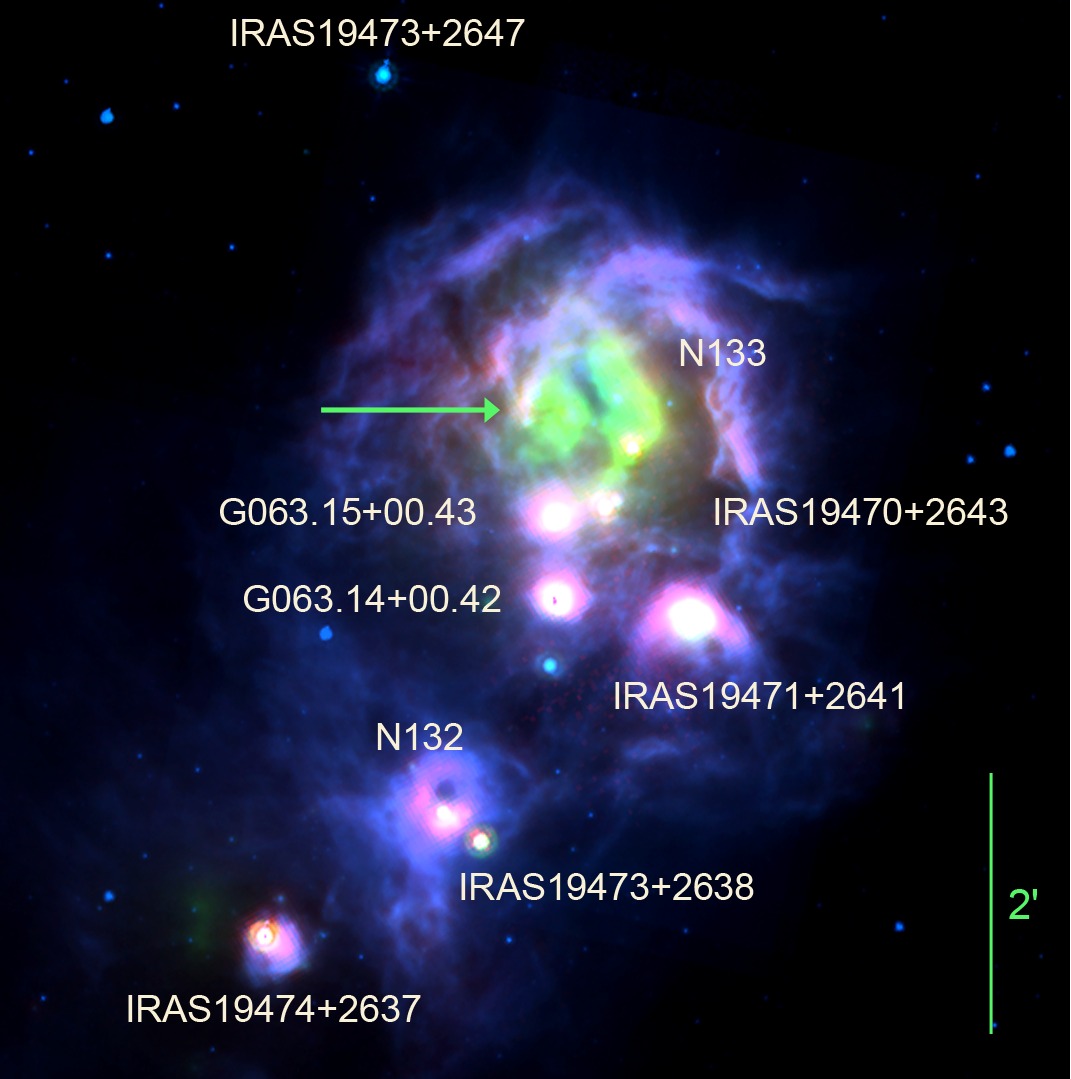}
  \caption{ Colour-composite image of the Sh2-90 complex centered at
 $\alpha_{2000} = 19^{\rm h}49^{\rm m}17.5^{\rm s}$,
$\delta_{2000} = +26^{\circ}49^{\prime}55^{\prime\prime}$. {\it Herschel}-PACS 70.0~$\mu$m
  (red), have been combined with {\it Spitzer}-MIPS 24.0~$\mu$m (green) and {\it Spitzer}-IRAC
   8.0~$\mu$m (blue) images. The arrow points to the 24.0~$\mu$m circular structure. 
  North is  up and east is  left.}
  \label{fig2}
\end{figure}

Figure 1 shows the morphology of the complex in optical (R-band) and IR (at 3.6 $\mu$m and 8.0 $\mu$m) bands. The 8.0 $\mu$m emission displays 
a central cavity
surrounded by a roughly thin annular shell emission.
The 8.0 $\mu$m
IRAC band contains emission bands at 7.7 $\mu$m and 8.6 $\mu$m, commonly attributed to polycyclic aromatic hydrocarbon (PAH) molecules; PAHs are believed to be destroyed in the ionized gas (Pavlyuchenkov et al. 2013), but are thought to be excited in the
 photo-dissociation region (PDR) at the interface of the \hii region and  molecular cloud by the absorption of far-UV photons  from 
 the exciting stars of the \hii regions.
The 8.0 $\mu$m IRAC diffuse emission is extended, well beyond the main shell, possibly because of the
leakage of UV photons through  holes into  the neutral material.
The emission at the 3.6 $\mu$m band is mostly from stellar sources, but this band also has
contributions from a weak, diffuse PAH feature at 3.3 $\mu$m.
Figure 2 displays the colour-composite image made with the {\it Spitzer}-IRAC 8.0~$\mu$m emission in blue, the {\it Spitzer}-MIPS
24.0~$\mu$m emission in green, and the {\it Herschel}-PACS 70.0~$\mu$m emission in red. Leaving the mid-IR (MIR) emission
from stellar and proto-stellar sources aside,  the  brightness distribution of the diffuse 24~$\mu$m emission differs from that of the diffuse 8.0~$\mu$m emission. 
The 24~$\mu$m emission is bright in the direction of the center of the \HII\ region, a region devoid of PAHs emission. As shown by the radiative 
transfer model of Pavlyuchenkov et al. (2013) this emission comes from very small dust grains (VSGs;  a few nm in size) located inside the ionized region, 
and out of thermal equilibrium after absorption of ionizing photons. Diffuse 24~$\mu$m emission is also observed in the direction of the neutral PDR 
surrounding the ionized region. Most of the diffuse 70~$\mu$m emission also comes from the PDR.  Compi{\'e}gne et al. (2010) have shown that 
stochastically heated VSG components can  contribute significantly (up to 50\%) to the diffuse emission at 70 $\mu$m.

 Figure 2  also displays bright compact dust structures at the locations of MSX and IRAS sources.  
It also shows one roughly circular  24 $\mu$m emission structure (pointed with an arrow), which lies close to a 250 $\mu$m clump (discussed in Sect. 6.4). 
These IR compact structures possibly represent the heated dust around  embedded massive source(s). We shall refer to the 
24 $\mu$m structure, the two MSX sources (G063.1549+00.4309 and G063.1422+00.4227), and  the IRAS source 
IRAS 19471+2641, as IR1, (IR2 and IR3), and IR4, respectively, in the text wherever necessary. 
We discuss these dusty structures in Sect. 6.4.


\begin{figure}[htp]
\centering {
\includegraphics[trim=0.0cm 1.2cm 0.0cm 0.7cm,clip=true,angle=0,width=8.0cm]{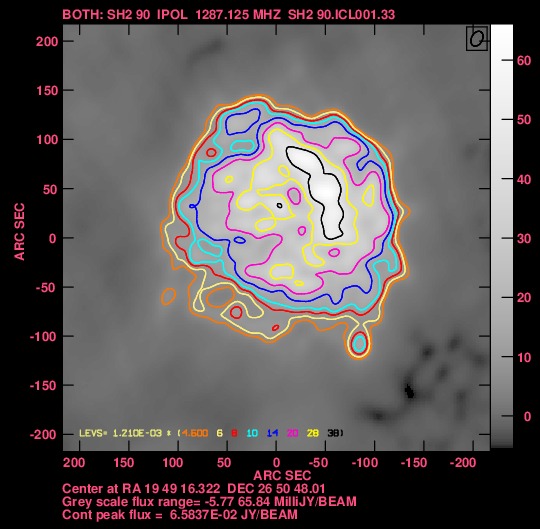}}
  \caption{Radio continuum map of Sh2-90  at 1280 MHz. 
   The contour levels are at 1.20 $\times$ (4.5, 6, 8, 10, 14, 20, 28, 38) \mjpb, where 1.21\mjpb~is the rms noise of the 
1280 MHz map. 
North is up and east is  left.}
\label{fig3}
\end{figure}
\begin{figure}[htp]
\center
  \includegraphics[trim=0cm 0.0cm 0cm 0.0cm,angle=0,width=8.0cm ]{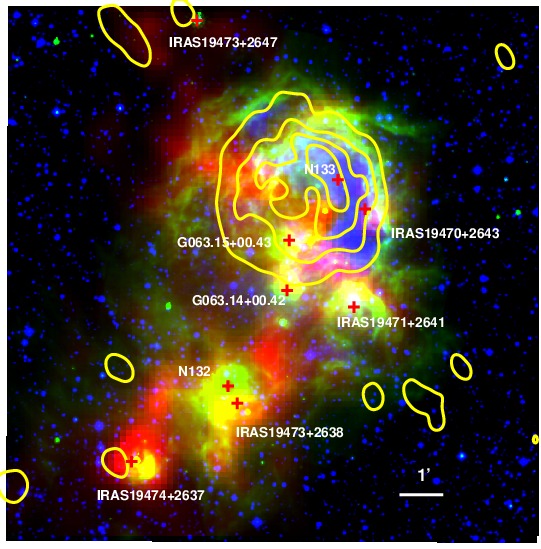}
  \caption{ Colour-composite image of the Sh2-90 complex centered at
 $\alpha_{2000} = 19^{\rm h}49^{\rm m}18^{\rm s}$,
$\delta_{2000} = +26^{\circ}49^{\prime}29^{\prime\prime}$. {\it Herschel}-SPIRE 350.0~$\mu$m
  (red), has been combined with {\it Spitzer}-IRAC 8.0~$\mu$m (green) and DSS2 R-band (blue) 
  images. The radio continuum 610 MHz contours are over plotted with yellow lines. The contour levels are at 
   7 $\times$ (3, 5, 9, 13) \mjpb, where $\sim$ 7 \mjpb~is the rms noise of 610 MHz map.
   North is  up and east is  left. }
  \label{fig4}
\end{figure}
Figure 3 shows the radio continuum view of the Sh2-90 \hii region at 23 cm.
This high-resolution (beam $\sim$ 16$\farcs$0 $\times$ 11$\farcs$0) map shows the non-uniform  distribution
of the ionized gas. The complex structure of the radio emission suggests that it could be due to
density inhomogeneities within the \hii region.
Our high-resolution 23 cm image allows us to identify a compact (diameter 0.6 pc) 
radio emission (at $\alpha_{2000}=19^{\rm h}49^{\rm m}09^{\rm s}$, $\delta_{2000}=+26^{\circ}48^{\prime}60^{\prime\prime}$) 
at the S-W edge of Sh2-90. The position of this compact radio source coincides with the location of 
IRAS 19471+2641 (marked in Fig. 3).

Figure 4 shows the colour-composite image in the R-band (blue), 8.0 $\mu$m (green), and 350 $\mu$m (red), over plotted with radio 
contours at 50 cm. The 50 cm image comprises a cometary head overplotted in the N-W direction and an intensity gradient  towards 
the S-E direction (see Fig. 4).
 In Fig. 4, the \hii region shows 
complex morphology in the optical, with diffuse, patchy, and irregular 
extended emission. An absorption lane in the optical band  is clearly seen 
at the center of the nebula, which is more prominent from the center to the N-E direction. The optical image
and radio contours display different  morphology at smaller scale. The N-W part of the nebula is bright in both the images. 
The striking difference  in Fig. 4 is the strong radio emission at the center and eastern side of the nebula, 
corresponding to the zone of optical absorption. This indicates  that the optical emission is absorbed by a dust cloud along  the line of sight. 
We observed cold neutral material,  prominent at longer wavelengths ($\geq$ 350 $\mu$m),  and 
distributed along an elongated structure which extends from the center to the  N-E  and S-E directions.
This implies that the cold gas is in front of the \hii region.  We note that although our low resolution 610 MHz map shows
the extended \hii region well, we do not see the compact radio source at the 3$\sigma$ upper limit level (i.e., $\sim$ 21 
mJy). The non-detection of the compact radio source  could be due to the fact that it is optically thick at 610 MHz. High-resolution
sensitive low-frequency maps would shed more light on the nature of this source.
\begin{figure*}
\centering{
  \includegraphics[trim=0cm 0cm 0cm 0cm, clip=true, angle=0,width=8.3cm ]{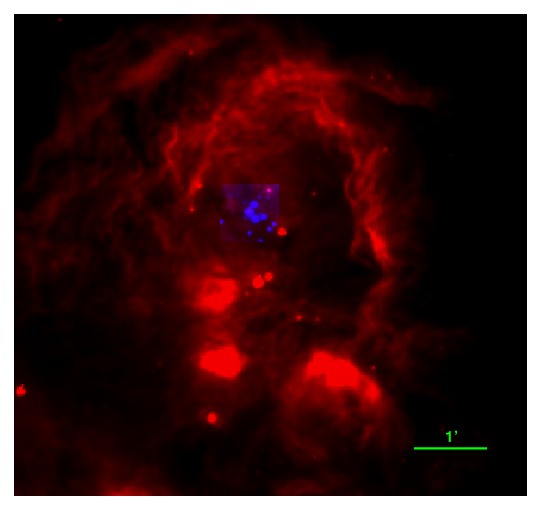}
  \includegraphics[trim=0cm 0cm 0cm 0cm, clip=true, angle=0,width=8.1cm ]{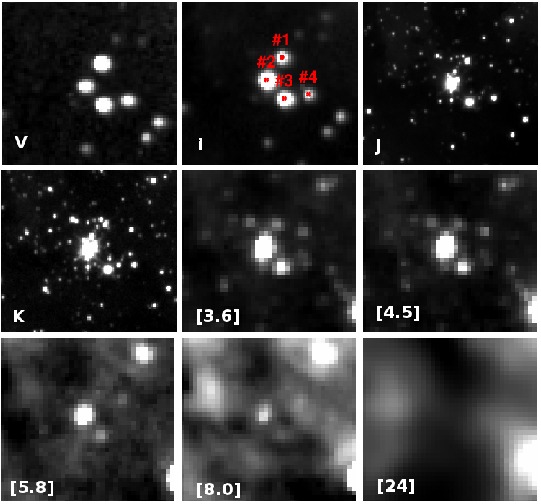}}
  \caption{{\it Left:} The   8.0~$\mu$m image (red; from {\it Spitzer}-IRAC) centered at
 $\alpha_{2000} = 19^{\rm h}49^{\rm m}13^{\rm s}$,
$\delta_{2000} = +26^{\circ}50^{\prime}42^{\prime\prime}$, showing structures  
  pointing to the  exciting star(s) candidates shown in R-band image (blue: from DSS2). {\it Right:}
Images showing  the  exciting star candidates (marked with  $\#1, \#2, \#3,$ and $\#4$) 
at various wavelengths  in the range 0.54 to 24 $\mu$m (i.e., images at V, I, J, K, 3.6 $\mu$m, 4.5 $\mu$m, 
5.8 $\mu$m, 8.0 $\mu$m and 24 $\mu$m bands).}
  \label{fig5}
\end{figure*}  

\subsection{Properties  of the ionized gas}
The flux densities (S$_\nu$) of Sh2-90, estimated by integrating over 4$\sigma$ contours yield S$_{23 cm}$ = 5.03 $\pm$0.5 Jy and S$_{50 cm}$ = 
3.20 $\pm$ 0.3 Jy, where $\sigma$ is the rms noise of the respective radio images. 
Our 23 cm flux density within uncertainty is close to the flux densities at 21 cm (5.8 $\pm$ 0.7 Jy) and 6 cm  (5.0 $\pm$ 1.0 Jy)
measured by Israel (1977). The flux densities at 23 cm, 21 cm, and 6 cm reflect a flat spectrum, indicating the region is optically thin at 23 cm.
Figure 3 shows that the radio emitting region  is almost circular in shape and has a diameter about 4$\farcm$8 (or 3.3 pc at 2.3 kpc). 
 Using the 23 cm flux density and assuming a spherically symmetric, optically thin homogeneous nebula, we determine the physical conditions and properties of the  ionized gas 
following the prescription given in Kurtz et al. (1994, and references therein). 
Considering   7370 K as the electron temperature ($T_{\rm e}$; Quireza et al. 2006) of the ionized gas, we derived  parameters 
such as the total number of Lyman continuum photons per second
($N_{Lyc}$ $\sim$ 22.5 $\times$ 1$0^{47}$ s$^{-1}$) 
coming from the associated massive star(s), the rms electron density ($n_e$ $\sim$ 144 cm$^{-3}$), 
the emission measure (EM $\sim$ 6.7 $\times$ 10$^{4}$ cm$^{-6}$ pc), and the mass of the ionized gas (M$_{ion}$ 
$\sim$ 55 \msun) for Sh2-90. 
The estimated $N_{Lyc}$ (Log ($N_{Lyc}$) $\simeq$ 48.35), suggests
that the spectral type of the ionizing source responsible for the ionization of Sh2-90
is an O8-O9V star according to Smith et al. (2002), or an O8-O8.5V star according to Martins et al. (2005). 
Because of our  limited knowledge on the dust  absorption of 
Lyc photons and  clumpy  nature of the medium,  the estimated $N_{Lyc}$ 
could be on the lower side.
The flux density of the compact \hii region
is $\sim$ 0.04 Jy.   Assuming that the compact \hii region associated with IRAS 19471+2641 is optically thin at 1280 MHz, we  estimated $N_{Lyc}$ as  $\sim$ 1.4 $\times$ 1$0^{46}$ s$^{-1}$ for an electron temperature $\sim$ 10000 K, which is equivalent 
to the Lyman continuum photons coming from a star of spectral type  B1 V (Smith et al. 2002).

\begin{table*}
\caption{Photometric magnitudes of candidate ionizing stars}
\centering
\begin{tabular}{cccccccccccc}
 \hline\hline
 ID  & RA  & Dec & V & I & J & H & K & [3.6] & [4.5] & [5.8] & 8.0]  \\
  & deg (J2000)   & deg (J2000) & mag & mag & mag & mag & mag & mag & mag & mag & mag  \\
  \hline
  $\#1$ & 297.31133 & 26.85463 & 16.50 & 14.92 & 13.92 & 13.38 & 13.19 & 12.63 & 12.24 & -- & --\\
  $\#2$ & 297.31265 & 26.85292 & 17.33 & 13.58 & 11.00 & 10.01 & 9.54 & 9.21 & 9.11 & 9.04 & 8.77\\
  $\#3$ & 297.31114 & 26.85151 & 16.59 & 14.08 & 12.47 & 11.63 & 11.30 & 11.15 & 11.11 & 10.90 & 10.44\\
  $\#4$ & 297.30907 & 26.85182 & 17.72 & 15.66 & 14.32 & 13.78 & 13.55 & 13.54 & 13.63 & -- & --\\
 \hline
\end{tabular}
\end{table*}

\subsection{Ionizing source(s) of Sh2-90}
The exciting star of Sh2-90 has not been  clearly identified.
According to Georgelin et al. (1975), the  exciting star of
Sh2-90 is an O9.5 III (ALS 10542; V=11.41 mag and B-V=0.78 mag) star, located at 
$\alpha_{2000}=19^{h}49^{m}14^{s}$, $\delta_{2000}=+26^{\circ}47^{\prime}39^{\prime\prime}$, thus $\sim$ 3$\farcm$5
 north of Sh2-90, i.e., significantly off center. 
Lafon et al. (1983), based on the
observed [O{\sc III}/H$_\beta$] line ratio  (by Chopinet \& Lortet-Zuckermann 1976), suggested that the ionizing star
should be of spectral type O8-O9 V, and more likely located inside the Sh2-90 region. 
In agreement with Lafon et al. (1983), our support for the exciting star being inside the bubble are as follows:

{\large $\star$} At 8.0 $\mu$m, the  region shows a cavity at its center, surrounded by PAH emission in the PDR (see Fig. 1). 

{\large $\star$} The 24 $\mu$m diffuse emission is strongest in the center of the nebula (see Fig. 2), 
implying that the dust heating source might be at the center of the nebula. This phenomenon has been 
observed in many bubbles associated with \hii regions (e.g., Deharveng et al. 2010)

{\large $\star$} Generally, the radio free-free emission 
comes from the immediate vicinity of the OB stars in young \hii regions (see Fig. 4). 

All this  evidence suggests that the exciting star(s) must be located inside the bubble. 
Thus, we searched for the exciting star(s) of the bubble within the dust cavity. First, we identified
all the luminous sources inside the bubble earlier than B2V stars. We followed the
same prescription as described in Samal et al. (2010), of rejecting the most-likely giants and
foreground sources based on $J$ vs. $J-H$ and $J-H$ vs. $H-K$ diagrams. After these eliminations we
remain with a massive O-type star located at the center of the bubble close to the strong 24 \mum emission. 
This source is associated with
a group of stars in its close vicinity, possibly part of an exciting cluster (Fig. 5, left). 
The  images of the ionizing candidates at various wavebands in the range 
0.54-24 \mum are shown in Fig. 5 (Right).  In optical bands only four sources are visible; their coordinates and  
fluxes at various wavelengths are given in Table 3. Our
high-resolution deep NIR images, however, reveal many faint sources, suggesting
the presence of a small cluster. The fainter sources  disappear
at IRAC-MIPS wavelengths possibly because of  a combined effect of lower sensitivity and poor resolution of IRAC-MIPS bands. 

 Out of four sources,  source $\#$2 ($J$=11.00 mag, $J-H$=0.99 mag) is the most luminous one.  
We assume that star $\#2$ is an O8-O9 MS star (using the prediction by Lafon et al. 1983); using its NIR photometry and synthetic colours of O stars by Martins \& Plez (2006),
its visual extinction is $\sim$ 10 mag (using the extinction law of Rieke \& Lebofsky 1985).
 Using the above extinction value, the observed J magnitude of the star and M$_J$-spectral type calibration table of Martins \& Plez (2006),
 we estimated that the distance to the star is in the range of 2.2-2.5 kpc.
This distance range is in agreement with the kinematic distance 2.1-2.4 kpc. 

Adopting  the distance of Sh2-90  in the range of 2.1-2.5 kpc, source $\#3$ ($J$ =12.47 mag, $J-H$= 0.84 mag) is consistent with a B2-B3 star, 
whereas  the nature of the other two neighboring sources, $\#1$ ($J$ =13.92 mag, $J-H$= 0.54 mag) and $\#4$ ($J$ = 14.32 mag, $J-H$ = 0.54 mag)  
appear to be low-mass stars. 
From the above discussion, we conclude that source $\#2$ is the most-likely ionizing star of Sh2-90. 
\section{Distribution of cold neutral material in the complex}
\begin{figure}
\center
  \includegraphics[trim=0cm 0cm 0cm 0cm, clip=true, angle=0,width=7.5cm ]{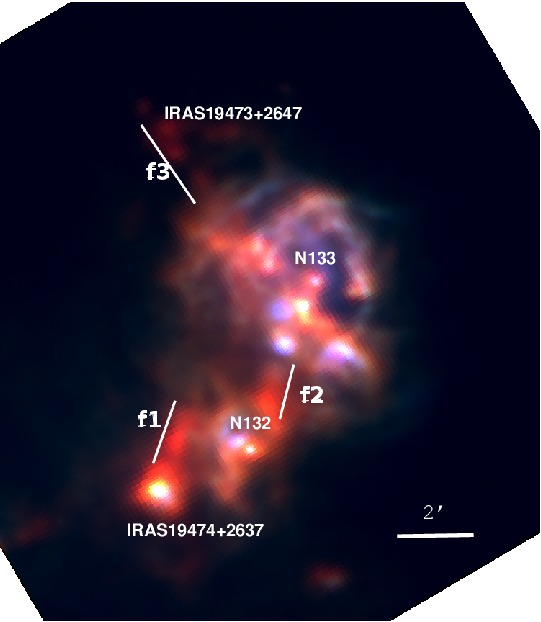}
  \caption{Colour-composite image of the Sh2-90 complex at {\it Herschel}  350 $\mu$m (red), 160 $\mu$m (green), and 70 $\mu$m (blue) 
bands centered at
 $\alpha_{2000} = 19^{\rm h}49^{\rm m}20^{\rm s}$,
$\delta_{2000} = +26^{\circ}50^{\prime}09^{\prime\prime}$.  The positions of the main star-forming sites are also marked (see Fig. 1). The solid lines represent the small filament-like structures (marked as f1, f2, and f3) of the region. North is up and east is left.}
  \label{fig6}
\end{figure}

\begin{figure}[htp]
\center
  \includegraphics[trim=0cm 0cm 0.0cm 0cm, clip=true, angle=0,width=8.0cm ]{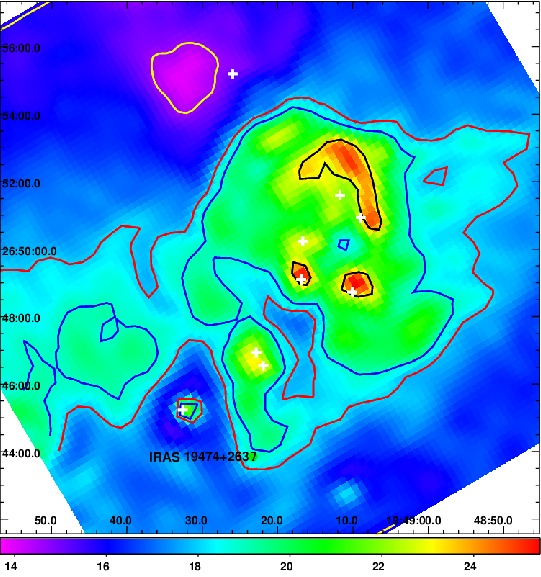}
  \includegraphics[trim=0.0cm 0cm 0.0cm 0cm, clip=true, angle=0,width=8.0cm ]{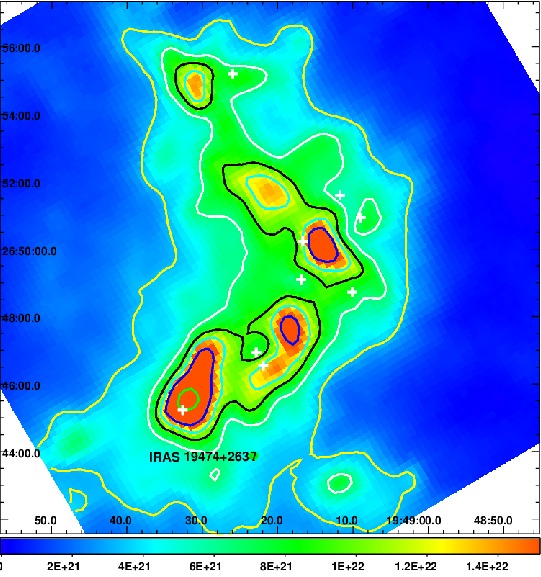}
  \caption{{\it Top:} Dust temperature map of the Sh2-90 complex centered at
 $\alpha_{2000} = 19^{\rm h}49^{\rm m}21^{\rm s}$,
$\delta_{2000} = +26^{\circ}50^{\prime}18^{\prime\prime}$. The horizontal colour bar is  labelled in unit of Kelvin (K) and the contours 
  are at 14.6 K (yellow), 18 K 
  (red), 19 K (blue) and 23 K (black). The map was obtained by fitting the dust emission between 160 $\mu$m and 500 $\mu$m. 
{\it Bottom:}  H$_2$ column density map of the same region. The horizontal colour bar is  labelled in units of cm$^{-2}$ and the  contours are at 3.3 (yellow), 6.4 (white), 
9.3 (black),  12 (cyan), 16 (blue) and 32 (green) $\times$ 10$^{21}$ cm$^{-2}$. North is up and east is  left.  
The labeled axes are in J2000 coordinates. The plus symbols represent 
the locations of the IR sources shown in Fig. 1.}
  \label{fig7}
\end{figure}

\subsection{Dust continuum distribution}
Recent {\it Herschel} observations have shown that SFRs are interconnected with network of filamentary structures (Arzoumanian et al. 2011), and 
that the massive-star-forming sites are  often associated with the main filament of the complex. In Fig. 6, we present  
a colour-composite image of the Sh2-90 complex at  350 \mum (red), 160 \mum (green), and 70 \mum (blue).
Because of the low resolution of the images, the individual components or cavities  seen at 8.0 $\mu$m (Fig. 1) are not 
clearly  distinguishable at 160 $\mu$m and 350 $\mu$m,  but the complex is more extended at {\it Herschel}-SPIRE wavelength
than at  8 $\mu$m,  showing new structures such as small filaments f1, f2, f3 ($\sim$ 0.7-1.3 pc in length and $\sim$ 0.3-0.4 pc in width) 
at the outer extent of Sh2-90. The fact that
these filaments are not observed at 70 \mum tells us that they are possibly far away from any source of strong radiation.
Filament-like structures are  often considered as the possible sites of on-going or 
future star formation.

\subsection{Physical conditions of the cold ISM}

Thermal emission from dust can be used to determine the physical conditions of a cloud such as
temperature, density, and mass. We derived the cold dust temperature using the {\it Herschel} fluxes.
We convolved all the {\it Herschel} maps to the  500 $\mu$m band resolution,
then all the maps are re-gridded to the pixel size of the 500 $\mu$m map. We then subtracted a background flux
(i.e., 14, 19, 18, 15, and 13 Jy/pix at 70, 160, 250, 350, and 500 $\mu$m, 
respectively), to minimize the contribution of excess emission along the line of sight. The background area was chosen far from the main cloud complex.
We fitted a modified black body of 
single temperature 
to the observed fluxes on pixel by pixel basis to construct the temperature map.

We used a dust spectral
index of $\beta$\,=\,2, and dust opacity law as given in Deharveng et al. (2012), in which the dust opacity per unit mass column density ($\kappa_\nu$) is
$\kappa_{\nu} = 10~(\nu/1000~{\rm GHz})^{\beta}$ cm$^2$/g. The choice to use the present opacity law
and its limitation  have been discussed in Deharveng et al. (2012). We adopted $\beta$\,=\,2, because this is close to the value found in
\hii environments (Anderson et al. 2012).
During the spectral fitting, we iteratively applied colour
correction factors by repeatedly fitting a temperature, and  then applying the corresponding correction
factors until successive fit results were unchanged. We used the  colour corrections given in  PACS calibration release note 
``PACS Photometer Pass-bands and Colour Correction Factors for Various Source SEDs''.
The colour-corrected temperatures are higher than those obtained without corrections by about 0.2 K to 0.7 K,
with a median 
difference  of $\sim$ 0.4 K.
From the temperature map, we then  derived the H$_2$ gas column density map using the dust continuum emission at 500 $\mu$m, gas-to-dust 
ratio R=100, and 2.8 as the mean molecular weight per H$_2$ molecule.

We derive the dust properties excluding and including 70 $\mu$m emission in the spectral fitting procedure. It has been demonstrated that a 
non-negligible fraction of the 70 $\mu$m emission comes from stochastically heated very small grains  (Compi{\`e}gne et al. 2010; Pavlyuchenkov et al. 2013); this is
especially true when an \hii region is present along the line of sight. 
In consequence, we prefer to consider the dust emission in the range from 160 $\mu$m to 500 $\mu$m.
Figure 7  shows the dust maps were obtained  excluding 70 $\mu$m emission in the spectral fitting procedure.
However, we also explored the dust properties including 70 $\mu$m emission in the spectral fitting. The inclusion of the 70 $\mu$m emission results in 
a temperature higher by a few Kelvin for the whole complex. 
In the following, while considering a specific structure we discuss the physical properties using the two temperatures (the physical properties that are obtained from including the 70 $\mu$m flux in the 
spectral fitting are given in the brackets).
 We note that if the  anti-correlation
between the dust temperature and the spectral index (i.e., $\beta$ $=$ (4.3 $\pm$ 0.5)$\times$ T$_d$$^{-0.2}$; Anderson et al. 2012) is valid, then the temperature of the cold regions (e.g., $<$ 15 K) in our map should be considered as an upper limit, and for 
warm regions (e.g., $>$ 30 K) it should  be treated as a lower limit.
The average temperature of the Sh2-90 complex is about   18.5 K (21.5 K); higher temperatures are found in PDRs and in some of the 
clumps. The temperature of the PDR associated with N133 is in the range of  22-24 K (23.5-27.5 K), whereas the temperature 
 of N132  is around 20 K (22.7 K). 
The PDR at the western side of Sh2-90 has a higher temperature and a
lower column density than that of its eastern edge, reflecting
the presence of warmer material at the western periphery of Sh2-90.
The Sh2-90 complex also 
contains a number of cool filament-like structures (see Fig. 6). 
The temperatures
in their directions lie in the range of  14-17 K (14.5-18 K).
The column density map  shows roughly an elongated interconnected  distribution of high column density ($>$ 6.4  $\times$ 10$^{21}$ cm$^{-2}$; white contour in Fig. 7, $bottom$) 
material, broadly running from S-E to N-W then towards N-E via Sh2-90, with local
maxima ($>$ 10$^{22}$ cm$^{-2}$; cyan contour in Fig. 7, $bottom$) at five locations. 

The regions of high column density correspond to 
low temperature zones and vice versa. The column density is 
highest in the direction of IRAS 19474+2637, reaching  a  value of $\sim$ 4.5 (3.4) $\times$ 10$^{22}$ cm$^{-2}$.

The column density map can be used to estimate the total  mass of the cloud using the relation:
\begin{equation}
M =  m_\mathrm{H}  \mu_\mathrm{H_2} \sum  N({\rm H}_2) A,
\end{equation}
where $m_\mathrm{H}$ is the hydrogen mass, A is  the  pixel area in cm$^{-2}$, and $\mu_\mathrm{H_2}$  is
the mean molecular weight per ${\rm H}_2$ molecule.
To estimate the mass of the cloud, we integrated over all the pixels in the column density map having a value $\geq$ 3.3  
$\times$ 10$^{21}$ cm$^{-2}$ (yellow contour in Fig.7, $bottom$). This value corresponds to 
four times  the background column density value on the map.   The 
resulting mass of the cloud is $\sim$  1.6  $\times$ $10^{4}$ $\msun$. This value is  lower by 20\% when 
derived from the column density map that uses a 70 $\mu$m flux in the process of spectral fitting. These values
are compatible
 with the mass $\sim$  4 $\times$ $10^{4}$ $\msun$ derived by Lafon et al. (1983) using CO observations.
 It seems that the Sh2-90  \hii 
region was born in a massive cloud of mass $\sim$ $10^{4}$ $\msun$ or greater. 

\subsection{Dust continuum clumps and swept up shell}
\begin{table*}
\centering
\caption{ Physical properties of the  clumps}
\begin{tabular}{|rrrrrrrr|}
\hline
\hline
  \multicolumn{1}{|c|}{ID} &
  \multicolumn{1}{c|}{$\alpha_{2000}$ (deg)} &
  \multicolumn{1}{c|}{$\delta_{2000}$(deg)} &
  \multicolumn{1}{c|}{$F_{\rm{250}}$ (Jy)} &
  \multicolumn{1}{c|}{d (pc)} &
  \multicolumn{1}{c|}{T$_{dust}$ (K)$\dag$} &
  \multicolumn{1}{c|}{M (\msun) $\dag$}&
  \multicolumn{1}{c|}{$n_{\mathrm{H}_2}$ (10$^4$ cm$^{-3}$)$\dag$} \\
\hline
 C1 & 297.383000 & 26.754361 & 149 & 0.64 & 18.9(19.5) & 206(186) & 1.8(1.7)\\
 C2 & 297.338125 & 26.772806 & 49 & 0.47 & 21.7(23.2) &  43(37) & 1.1(0.9)\\
  C3 & 297.289667 &  26.817600 & 55 & 0.55 & 23.9(26.9) & 37(28) & 0.5(0.4)\\
  C4 & 297.319542 & 26.821258 & 40 & 0.46 & 24.0(26.9) & 26(20) & 0.6(0.5)\\
  C5 & 297.310208 & 26.836147 & 132 & 0.69 & 21.4(25.8) & 125(75) & 0.9(0.6)\\
  C6 & 297.334583 & 26.862675 & 34 & 0.42 & 21.8(27.6) & 31(17) &  0.9(0.6)\\
  C7 & 297.304542 & 26.847822 & 98 & 0.79 & 21.2(26.2) & 95(89) &  0.5(0.3)\\
  C8 & 297.306375 & 26.884567 & 14 & 0.31 & 23.8 (25.4) & 10(8) &  0.9(0.7)\\
  C9 & 297.282208 & 26.849556 & 27 & 0.43 & 23.9 (25.3) & 18(16) &  0.5(0.5)\\
\hline
\end{tabular}
\flushleft\small{ $\dag$ The values in brackets are estimated including the 70 $\mu$m emission in the temperature determination.}
\end{table*}
\begin{figure}[t]
\center
  \includegraphics[trim=0cm 0cm 0cm 0cm, clip=true, angle=0,width=9cm, height=10cm ]{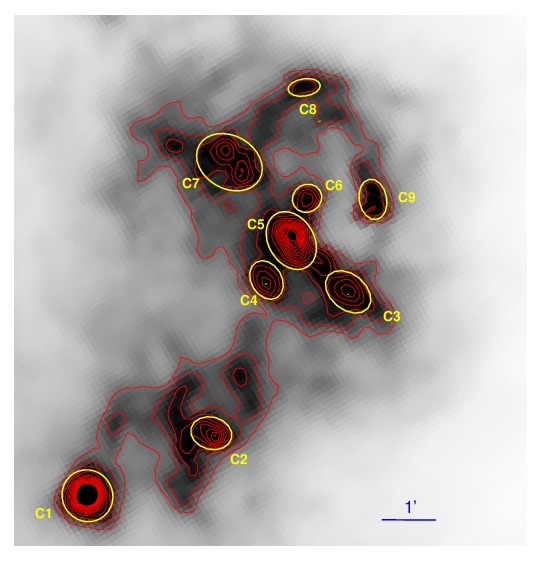}
  \caption{ The  250 $\mu$m cold dust emission contours superimposed on the 250 $\mu$m image. The image is centered at
 $\alpha_{2000} = 19^{\rm h}49^{\rm m}16^{\rm s}$,
$\delta_{2000} = +26^{\circ}49^{\prime}39^{\prime\prime}$. The contour levels are in the range
 from  980 MJy sr$^{-1}$   to 3000 MJy sr$^{-1}$ at intervals of 199 MJy sr$^{-1}$. 
 North is  up and east is left. The clumps discussed in the text are marked  C1 to C9. The yellow ellipses are the
apertures used to integrate the 250 $\mu$m fluxes. }
  \label{fig8}
\end{figure}
Figure 8  shows the 250 $\mu$m  emission contours of the complex.  The 250 \mum emission map reveals  nine compact 
structures associated with the main star-forming sites, 
and are marked in the figure. 
To estimate the mass of these compact structures, we  integrated their 250 \mum fluxes above the local background using
elliptical apertures. Since the clumps are part of interconnected extended structures, a clear boundary
cannot  be assigned to them. We chose aperture (boundary)  such that it closely encloses the flux within the 
outer contour level,  and  separated the clumps from the extended structures to make them as 
single  entities (see Fig. 8).
Since we  need to know the temperature to estimate the mass, 
we measured the average temperature of each compact structure from the temperature maps using the
same elliptical aperture that was used for the flux estimation. We also subtracted  
background flux from  each clump to minimize the excess emission along the line of sight plus the contribution 
of the diffuse cloud in which they are embedded. The background areas were chosen close to each compact structure;
however, since the background level is non-uniform even on a small scale, its accurate measurement 
is always problematic, which adds an extra uncertainty in the 
real flux estimations of the compact structures. Keeping all these uncertainties in mind, we derived 
the mass (gas $+$ dust) of the compact structures using the  relation (derived from Hildebrand 1983) for optically thin emission,
\begin{equation}
M = 100\,\,\frac{F_{\rm{\nu}}\,\,
D^2}{\kappa_{\rm{\nu}}\,\,
B_{\rm{\nu}}(T_{\rm{dust}})},
\end{equation}
where $F_{\rm{\nu}}$ is the measured integrated flux density, $D$ is
the distance of the source, $\kappa_{\mathrm{\nu}}$ is
the dust opacity per unit mass at frequency ${\rm{\nu}}$ (see Sect. 5.2),
and $B_{\rm{\nu}}(T_{\rm{dust}})$
is the Planck function for a dust temperature $T_{\rm{dust}}$.
We have assumed a gas-to-dust ratio of 100. A requisite of this formulation is
that the compact structures should be optically thin at the adopted frequency. Assuming the same kind of dust discussed in Section 5.2, an  optical 
depth $\sim$ 1 at 250 $\mu$m corresponds to a column density value  $\sim$ 1.5 $\times$ 10$^{24}$ cm$^{-2}$. The maximum
column density (i.e., $\sim$ 4.5(3.4) $\times$ 10$^{22}$ cm$^{-2}$; see Sect. 5.2) found in the direction of Sh2-90 is 
 significantly lower than the value $\sim$ 1.5 $\times$ 10$^{24}$ cm$^{-2}$; thus, we considered that all our compact structures are optically 
thin at 250 $\mu$m.

We give the peak position, integrated flux density at 250 $\mu$m ($F_{\rm{250}}$), mean diameter (d; the geometric mean of semi-major and semi-minor axes of ellipse),
measured dust temperature ($T_{\rm{dust}}$), mass ($M$), and  density ($n_{\mathrm{H}_2}$) of  each compact structure in Table 4.
We note $n_{\mathrm{H}_2}$ has been estimated assuming a uniform density inside the aperture.  Bergin \& Tafalla (2007) defined a clump 
as having a size 0.3-3 pc and containing 50-500 \msun~ and 
defined a core as being around 0.03-0.2 pc and having a mass of 0.5-5 \msun.  We observed that the size, mass, and density of the clumps  
are in the range 0.3-0.6 pc, 10(8)-206(186) \msuns, and 0.5(0.4)-1.8(1.7) $\times$ 10$^{4}$ cm$^{-3}$, respectively. Here again the values in
brackets are obtained including the 70 $\mu$m emission in the temperature determination.
Following the nomenclature used by Bergin \& Tafalla (2007), in the following we refer to these compact structures simply as clumps.

To test whether these clumps are gravitationally
bound entities or if some of them are unbound  transient structures, we
estimated the Bonnor-Ebert critical  mass ($M_\mathrm{BE}$) using the following relation from Lada et al. (2008):
\begin{align*}
        M_\mathrm{BE} &\sim 1.82\,\left(\frac{n_{\mathrm{H}_2}}{[10^4]{cm^{-3}}}\right)^{-1/2}\,\left(\frac{T}{[10]{K}}\right)^{3/2}\,\msun.
\end{align*}
Within the framework of this simple approach, the clump with mass greater than $M_\mathrm{BE}$ will collapse under the effect of self-gravity in the
absence of other forces, whereas the  clump  with mass less than $M_\mathrm{BE}$ is not gravitationally bound or unstable.
To derive $M_\mathrm{BE}$, we used  the  temperature and density given in Table 4. The results are such that all the
clumps seem to be gravitationally bound and so can lead to star
formation, except clumps C8 and C9, which are marginally bound.

Kauffmann \& Pillai (2010) suggested an approximate threshold for massive star (M $>$ 10 \msun) formation by comparing
clouds with and without massive star(s). The clouds expected to form  massive stars obey typical
mass-size relation of the form: m(r) $>$ 870 $\msun$ $\times$ (r/pc)$^{1.33}$,
where r is the effective radius in pc.
 The clumps C1, C2, C3, C4, C5, C6, and C7 are already associated with intermediate to massive
stars (discussed in Sect. 6.4), suggesting that they have already  collapsed, although the 250 $\mu$m dust peak in some of the
clumps (C3, C4, and C7) is not co-spatial with the exact position of their massive member(s). 
This could be dbecause they are the remnants of 
the original clumps. We do not favor that they really represent new sites of star formation because 
these clumps are warmer in our temperature map. Clumps C8 and C9 do not seem to be associated with any 
active star formation (see Sect. 6.4).

The large-scale distribution of the 250 $\mu$m cold dust emission at the periphery of Sh2-90 broadly represents matter accumulated during the 
expansion of the \hii region (for example see Fig. 1 of Deharveng et al. 2010), which now resides in a shell. To estimate the mass of the
accumulated matter, we integrated the 250 $\mu$m emission over a contiguous region  in an irregular aperture that closely encloses 
the material in the shell.   Using the average dust temperature $\sim$23 K (25.5 K) of the shell,
we estimated the mass of its molecular content  as $\sim$ 800 (610) \msun. Considering the ionized 
mass $\sim$ 55 $\msun$ (see Sect. 4.3) plus the molecular mass $\sim$ 800 \msuns (see Sect. 4.3), the total mass 
content of the  region is $\sim$ 855 (665) \msun. Assuming that this 
total mass was distributed homogeneously in a sphere of 
radius 1.3 pc (the mean radius of the shell), we estimated the volume average density of 
the original medium as $\sim$  2.6(2.0) $\times$ 10$^3$ cm$^{-3}$. 


\section{Young stellar populations of the complex and their nature}
\begin{figure}
\center
 \includegraphics[trim=0.1cm 0.1cm 0.1cm 0.1cm, clip=true,angle=0,width=5.0cm ]{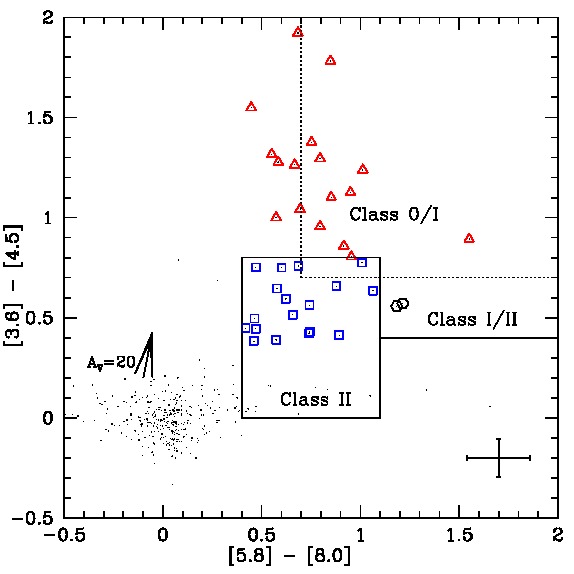}
 \includegraphics[trim=0.1cm 0.1cm 0.1cm 0.1cm, clip=true,angle=0,width=5.0cm ]{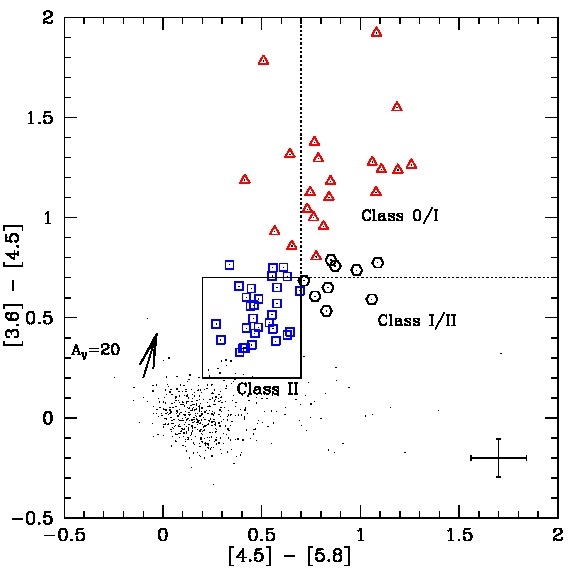}
 \includegraphics[trim=0.1cm 0.1cm 0.1cm 0.1cm, clip=true,angle=0,width=5.0cm ]{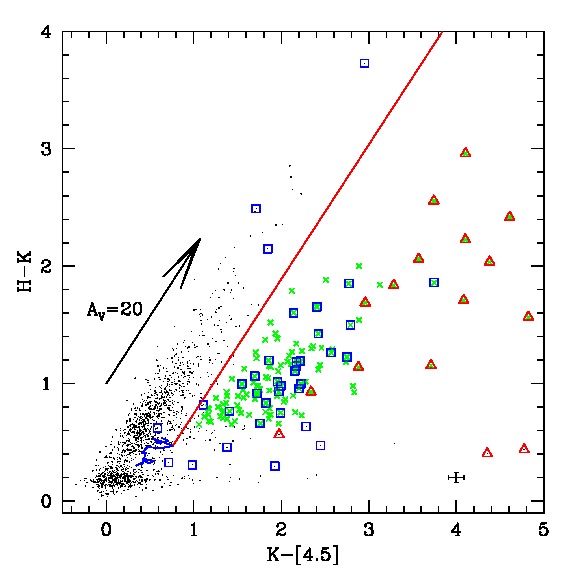}
  \caption{ {\it Top:} IRAC [3.6]-[4.5] vs. [5.8]-[8.0] CC diagram with boxes representing the
boundaries of different classes of sources.  {\it Middle:} IRAC [4.5]-[5.8] vs. [3.6]-[4.5] CC diagram
with boxes representing the boundaries of different classes of sources.
The Class 0/I, Class I/II, and Class II sources are marked with
triangles, hexagons, and squares, respectively.
{\it Bottom:}
The H-K vs. K-[4.5] CC diagram. The curved solid line (blue) is the MS locus of late M-type dwarfs (Patten et al. 2006). The long solid line (red) represents the reddening vector
from the tip of a M6 dwarf. 
The crosses represent the extra IR-excess sources identified from this diagram, whereas the YSOs identified only with the
{\it Spitzer} bands are marked in squares (Class I/II plus Class II) and triangles (Class 0/I), respectively.
A reddening vector of A$_V$ = 20 mag and mean error bars of the colours are shown in these diagrams. 
The mean colour error of [3.6]-[4.5], [4.5]-[5.8], [5.8]-[8.0], K-[4.5], and H-K are 0.09, 0.14, 0.16, 0.08, and 0.04,
respectively.}
 \label{fig9}
\end{figure}
\subsection {Identification and classification of YSOs}

The circumstellar emission from the disk and envelope  in the case of YSOs dominates at long wavelengths (in near to far-IR),
where the spectral energy distribution (SED) significantly
deviates from the pure photospheric emission. 
Here, we summarize the  method that we adopted to 
identify and classify YSOs  using our multi-band photometric data set.

First,  we  match the  GLIMPSE catalog (Benjamin et al. 2003) at
3.6, 4.5, 5.8, and 8.0 $\mu$m bands  with our deep NIR catalog 
using a 1$\farcs$2 radial matching tolerance. 
We then identified and classified the Class I 
(protostars with in-falling envelopes, including flat spectrum objects), Class II 
(pre-main-sequence (PMS) stars with optically thick disks), and Class I/II (sources that display characteristics of both Class I and Class II)  YSOs 
using  [3.6]-[4.5] vs. [5.8]-[8.0] (Allen et al. 2004) and [3.6]-[4.5] vs. [4.5]-[5.8] (Hartmann et al. 2005) colour-colour (CC) diagrams. The details about these diagrams 
can be found in Samal et al. (2012) and references therein. These diagrams are shown in Fig. 9 ($top$ and $middle$ panels). 
In these figures the zones of  Class 0/I, Class I/II and Class II  sources are marked with lines, whereas the 
foreground, MS, and Class III objects are generally found  around (0,0). We note that the classifications of sources near the boundaries of respective zones
using such diagrams are always tentative.  In order to constrain the contamination of non-YSO candidates to our sample, 
we analyzed a control field (at $\alpha_{2000} = 19^{h}49^{m}05^{s}$, $\delta_{2000} = 27^{\circ}44^{\prime}49^{\prime\prime}$) as of equal area 
to the target field, located approximately 30$\farcm$0 away from the Sh2-90 region. 
We constructed  the same CC plots (not shown) for the control region\footnote{The control region data were obtained from the GLIMPSE
archive (Benjamin et al. 2003) and UKIDSS Galactic Plane Survey (GPS; Lawrence et al. 2007) at NIR bands.} 
The control field gives statistical distribution of non-YSO sources (including reddened background sources and scattered field distribution) in 
the same Galactic direction as of the cluster region along the line of sight.
 In order to avoid such non-YSO candidates, we selected YSOs in the cluster region after applying colour cuts. The colour cuts were chosen on the basis of
 distribution of  point sources in the CC plot of the control field. In this approach we first selected YSOs from the [3.6]-[4.5] vs. [5.8]-[8.0] 
diagram (marked with triangles, hexagons, and squares for Class 0/I, Class I/II, and Class II, respectively),  we then overplotted 
these YSOs in the [3.6]-[4.5] vs. [4.5]-[5.8] diagram. We found that some of the Class II sources identified in the [3.6]-[4.5] vs. [5.8]-[8.0] diagram 
fall in the Class I  and Class I/II zones of [3.6]-[4.5] vs. [4.5]-[5.8] diagram, possibly due to the effect of reddening. This led us to slightly 
modify the classification boundary of the [3.6]-[4.5] vs. [4.5]-[5.8] diagram. After a minor modification, we selected Class 0/I, Class I/II, and 
Class II  YSOs from the [3.6]-[4.5] vs. [4.5]-[5.8] diagram, which are marked with the same symbols as in Fig. 9 ($top$). Since the majority of the 
Class II YSOs of the [3.6]-[4.5] vs. [5.8]-[8.0] CC diagram fall in the Class I/II zone of the [3.6]-[4.5] vs. [4.5]-[5.8] diagram and vice versa, we 
therefore tentatively consider all the Class I/II YSOs as Class II YSOs.

For sources that are not detected  in the [5.8] and/or [8.0]  bands, we use the H-K vs. [K]-[4.5] CC diagram to identify 
extra YSOs (shown in Fig. 9, $bottom$).  This diagram basically recovers  YSOs, 
which are not detected at longer wavelengths because of high background level in the \hii region environments.
In this diagram,  the sources located to the right of the MS reddening vector are likely to be YSOs with NIR excess.
We minimize the contamination of other sources to this diagram  
by selecting sources with [3.6] mag $<$ 14.5 (Fazio et al. 2004), and  applying a cut in H-K colour (i.e., 0.65 mag). The colour cut was chosen 
by comparing the distribution of already IRAC classified YSOs in Fig. 9 ($bottom$) to
avoid  scattered and reddened field distribution that we noticed on the lower right side of the MS reddening vector 
in the CC diagram of control field. We then considered only those sources as NIR-excess candidates, whose excess is more than  
1$\sigma$ (where, $\sigma$ is the mean colour error) from the MS reddening line.  In this approach 
we may miss a few YSOs, but the selected candidates would be more reliable sources with disks and envelopes.  In Fig. 9, the already identified 
Class I and Class II  YSOs are shown in triangles and squares, respectively. The crosses that are not surrounded by triangles or squares, 
are the additional NIR-excess YSOs (i.e.,  sources not identified as YSOs based on IRAC colours). They  are possibly Class II 
YSOs, as their positions fall in the regime
of the already classified IRAC Class II sources; however, we termed them as NIR-excess YSOs in the
present work. Finally, we visually inspected the counterparts of all the YSOs in our high-resolution \jhk images to reject
the most-likely unresolved extended sources. However, if we accept that the number of  non-YSO objects still lying in  our selected YSO zones in the 
CC plots of the control field, then this suggests that our YSO sample is likely to be contaminated by less than $10\%$.

In total, we have identified 129 sources in the Sh2-90 complex with excess IR emission.  Of
these, 21 have excess consistent with Class I YSOs, 34 have excess consistent with
Class II or Class I/II YSOs, and 74 are termed as NIR-excess YSOs (i.e., sources not classified as YSOs based on IRAC colours). 
We did not classify diskless YSOs (Class III sources),
because with the present data sets they are indistinguishable from field stars. We note that although the NIR and IRAC colours are very useful 
for identifying YSOs,  the YSOs classification can be altered based on the high degree of 
reddening and viewing angle. Nevertheless, the assumed classification scheme provides a good
representation of the respective YSO classes.  The catalogue of the identified
YSOs in the present analysis is given in Table 5. 
 A sample of  Table 5 is given here; the complete
table is available in electronic form at the CDS
via anonymous ftp to cdsarc.u-strasbg.fr (130.79.128.5) or via
http://cdsweb.u-strasbg.fr/cgi-bin/gcat?J/A+A.
\begin{figure}[btp]
\center
 \includegraphics[angle=0, width=8.0cm ]{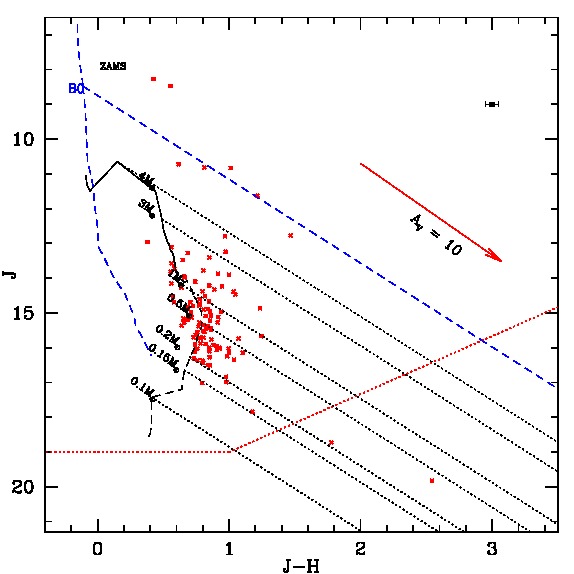}
  \caption{ $J$ vs. $J-H$ CM diagram for YSOs of the Sh2-90 complex.
The PMS isochrones of 1 Myr from Siess et al. (2000) and Baraffe et al. (1998)
are drawn in solid and dashed curved lines (black), respectively,
for a distance of  2.0 kpc and zero reddening. The reddening vectors
corresponding to 0.1, 0.15, 0.2, 0.5, 1, 3, and 4 $\msun$  are drawn in
dotted slanting lines. The ZAMS (vertical dashed line in blue), along with the
reddening vector (slanted line in blue) from the tip of B0 star, is also shown.
The average error in colour is shown on the upper-right side of the figure.
}
 \label{fig10}
\end{figure}
\subsection {Mass distribution of the YSOs}
Having identified  YSOs, we can quantitatively constrain their stellar masses with an assumed age. Since YSOs generally show excess emission at long wavelengths, to minimize the effect of excess emission on masses we use $J$ vs. $J-H$ colour-magnitude (CM) diagrams.
 Figure 10 represents the intrinsic  $J$ vs. $J-H$ CM diagram for  109 sources (out of the 129 excess sources), having counterparts 
 in $J$ and $H$ bands. To produce the intrinsic CM diagram, the extinction in front
of YSOs is derived by tracing back their observed colours
to the CTT locus or its extension in ($H- K$) vs. ($J-H$) CC
diagram  along the reddening vector. The solid and dashed curves in the figure denote the loci of 1 Myr
PMS isochrones by Siess et al. (2000) for 1.2 $M_\odot$ $\leq$ $M$ $\leq$  7 $M_\odot$ and 
Baraffe et al. (1998) for 0.05 $M_\odot$ $\leq$ $M$ $\leq$ 1.4 $M_\odot$, respectively. The
dotted slanting
lines are the reddening vectors for  4, 3, 1, 0.5, 0.2, 0.15, and 0.1 $M_\odot$ stars for
1 Myr isochrones.  Figure 10 suggests that the ages of most of the YSOs are probably 1 Myr or less.
Here, we would like to mention that age estimation of young clusters  by comparing the observations with the theoretical isochrones 
is prone to several errors such as unknown extinction, excess emission due to disk, variability, and binarity; the ages can be therefore highly discrepant. 
Here, we do not intend to estimate the exact masses of the 
YSOs, rather we would like  to know their approximate masses. Since the PMS isochrones for 
low-mass stars are very close to each other, a change in age of 0.5-1 Myr would not 
change drastically the masses of the low-mass YSOs. Thus, we considered 1 Myr as a representative
age to estimate the mass of the YSOs.
It is worth noting that adopting different sets of evolutionary tracks
would provide different values of stellar masses. However, 
for low-mass objects, the tracks of Siess et al. (2000) are close  
to those of Baraffe et al. (1998). The agreement between ages and masses of these two models is 
within 20-40\%.

The  dotted (red) line indicates the approximate completeness limit of our NIR data. 
 Figure 10 shows that for the  assumed age of 1 Myr, the lowest
YSO mass down to which our J-band image is {\bf complete} about 0.2 \msun~ for \av $\sim$ 8 mag (the mean \av of 
all the YSOs). From the figure it appears that most of the identified YSOs have 
masses  in the range of 0.2 - 3 $M_\odot$, except a few sources whose locations fall close to the reddened 
zero-age-main-sequence (ZAMS) star of spectral type B0.

\subsection{The spatial distribution of YSOs}
\begin{figure}
\center
 \includegraphics[width=9.0cm ]{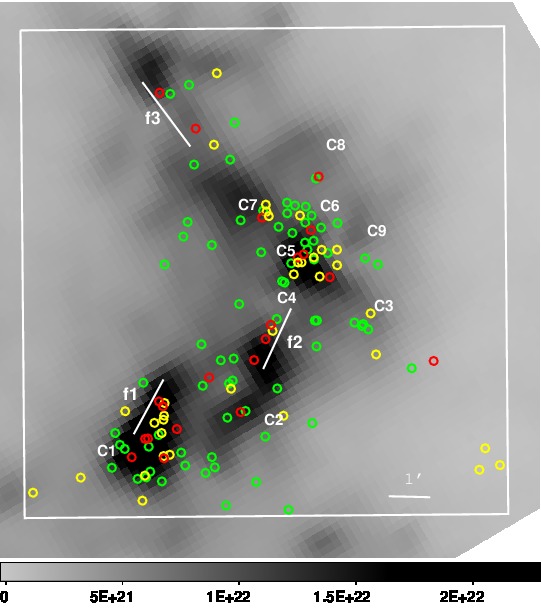}
  \caption{Spatial distributions of Class I (red circles), Class II (yellow circles) and NIR-excess (green circles) YSOs  
 on the column density map centerd at
 $\alpha_{2000} = 19^{\rm h}49^{\rm m}22^{\rm s}$,
$\delta_{2000} = +26^{\circ}50^{\prime}15^{\prime\prime}$. The horizontal  bar is  labelled in units of  cm$^{-2}$. The box represents the area for which
Class I, Class II  and NIR-excess YSOs have been searched.
   North is up and east is left}
  \label{fig11}
\end{figure}
We identified 129 YSOs using IRAC and NIR colours. The spatial distribution of these YSOs on the
column density map is shown in Fig. 11.  The YSOs (Class 0/I in red, Class II in yellow, and NIR-excess in
green) are found to be preferentially located in/around regions of high column density, with enhanced concentration at the locations
of C1 and C5. We note that the location of C5 is close to the \hii region Sh2-90, whereas C1 is located far away 
from it. We discuss possible star formation processes at these locations in Sect. 8.
The figure also displays that the majority of the  Class 0/I YSOs are found to be distributed in the directions of  
f1, f2, f3, and clump C5, whereas the NIR-excess YSOs show a slightly scattered distribution,
but in general the YSO distribution follows the elongated nature of original cloud seen in the CO map of Lafon et al. (1983) 
and in our 350 $\mu$m image (see Figs. 4 and 6). 
\begin{figure*}
\centering{
  \includegraphics[trim=0cm 0cm 0cm 0.0cm, clip=true, angle=0,width=15cm ]{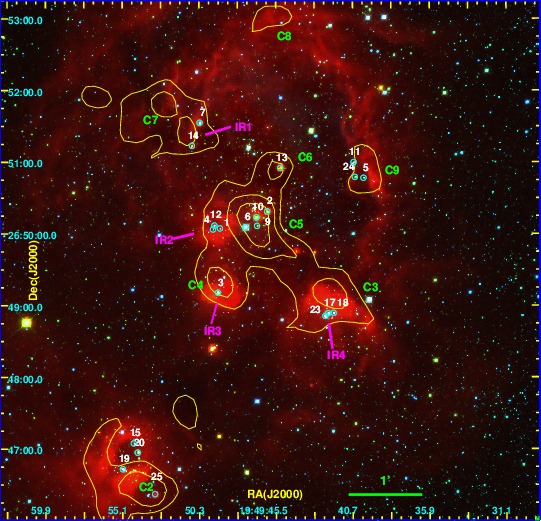}}
  \caption{{\it Left:} The colour-composite image of  Sh-90 centerd at
 $\alpha_{2000} = 19^{\rm h}49^{\rm m}13^{\rm s}$
$\delta_{2000} = +26^{\circ}49^{\prime}54^{\prime\prime}$, showing point sources at 1.25 $\mu$m (blue), 2.14 $\mu$m (green), and
 5.8 $\mu$m (red). The names refer to the different IR blobs (see Sect. 6.4) and 250 \mum clumps (see Sect. 5.3) 
identified in the complex. The yellow contours represent the 250 $\mu$m clumps. The contours are at 1450, 2000, and 3000 MJy sr$^{-1}$. 
The small circles mark the position of the NIR sources that lie within these IR blobs and clumps. North is up and east is to the left. }
\label{fig12}
\end{figure*}
To verify  whether the observed trend of the YSO distribution in the small filament-like structures (f1, f2, and f3) 
is due to the effect of extinction to the photometric colours or if they are really  sources with emission from 
the  disks and envelopes, we estimated the column density towards these structures and relate its effect on the photometric colours. 
The mean column density of  f1, f2 and f3, is $\sim$ 1.4 $\times$ 10$^{22}$ \cmsq, $\sim$ 1.3 $\times$ 10$^{22}$ \cmsq, and $\sim$ 8.9 $\times$ 10$^{21}$ \cmsq,
respectively, although the column density in the direction of individual YSO can vary depending on 
their exact location. These values correspond to mean visual extinction $\sim$ 15, 14, and 9.5 mag, respectively
(using $N(\mathrm H_2)$ = 0.94 $\times$ 10 $^{21}$  \av cm$^{-2}$  mag$^{-1}$; Bholin et al. 1978 and Rieke \& Lebofsky 1985).
These extinction values are possibly underestimated owing to the low spatial  resolution of 
our column  density map. However, a visual extinction value of 20 mag  can only produce a shift of 1.07, 0.26, 0.06, and 0.05 mag in
K-[4.5], [3.6]-[4.5], [4.5]-[5.8], and [5.8]-[8.0] colours, respectively (based on extinction laws of Rieke \& Lebofsky 1985 and Flaherty et al. 2007). 
Thus, we anticipate that the most of the identified YSOs are  real YSOs of 
the complex. 
 We note that the quoted extinction 
 values are  determined assuming that the extinction law of the general diffuse ISM 
 (i.e., total-to-selective extinction=3.1) is valid,  which may not be the case for very dense regions.  
  In such cases, the extinction values can be increased by  a factor of 1.37, 
 if \rv  reaches  a value of 5.5 (see Weingartner \& Draine 2001; Dunham et al. 2011).

\subsection{Nature of sources within IR blobs and/or clumps}
Polycyclic aromatic hydrocarbon emissions can be used as tracers of  embedded B-type star formation (Peeters 2004). These stars have the ability to  heat the surrounding dust to high 
temperatures, and can excite the PAH bands and fine-structure lines.   We see several compact  dust emission features 
(e.g., IR1, IR2, IR3, and IR4) at 8 $\mu$m, 24 $\mu$m, and/or 70 $\mu$m (see discussion in Sect. 4.1 and Fig. 12), similar to those seen at the peripheries 
of the bubbles such as RCW 79 (Zavagno et al. 2006; their clumps $\#2$ and $\#4$) 
and RCW 120 (Zavagno et al. 2007; their clump $\#4$).
These compact IR structures are possibly a tracer of low luminosity (Log (L/\lsun)=1.5-4.4)  embedded massive B-type stars. 
Hence, for  discussion purposes, hereafter we collectively  these compact IR structures as IR blobs.  

In order to identify the probable massive members in the IR-blobs
and  in the clumps C1 to C9 (some of them are co-spatial with the IR-blobs), we  
used our sensitive, high-resolution  $JHK_{\rm s}$ catalog. We  visually searched for the counterparts of the 
bright IRAC and NIR sources, within the approximate boundary of these blobs or clumps. 
The  probable sources are  marked in Fig. 12, and the positions of these sources 
are also shown in the $(J-H)$ vs. $(H-K)$ (Fig. 13, $right$) and $J$ vs. $(J-H)$ (Fig. 13, 
$left$) diagrams. Many sources show NIR excess in Fig. 13 ($right$); therefore, the 
stellar luminosity of such sources in the $J$ vs. $(J-H)$  diagram is uncertain and should be 
considered as an upper limit. 
\begin{figure*}[htp]
\centering{
  \includegraphics[angle=0,width=6.5cm ]{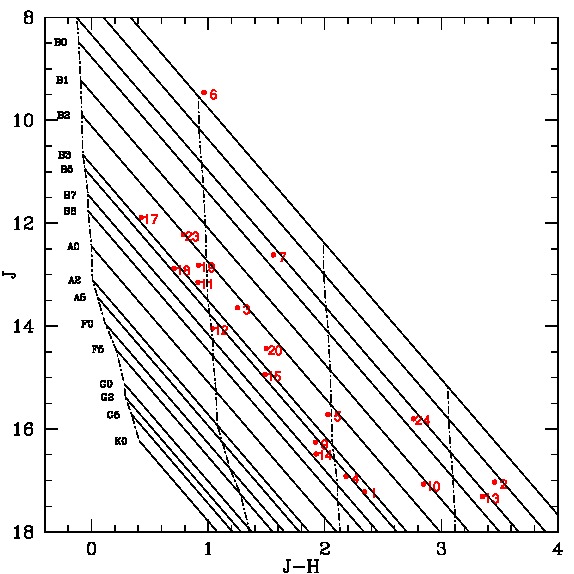}
  \includegraphics[angle=0,width=7.0cm ]{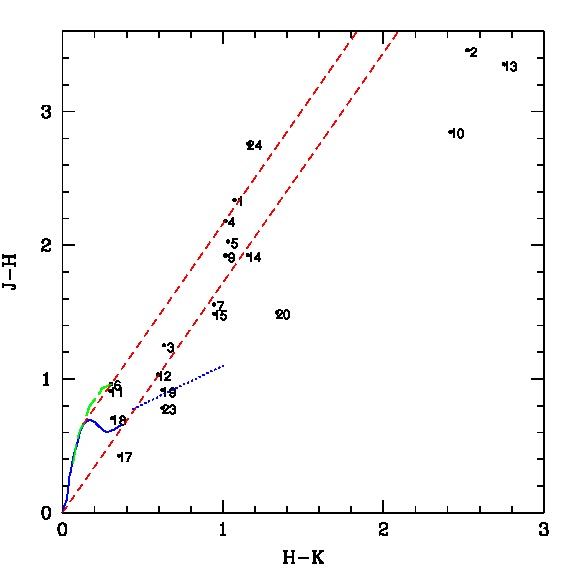}}
  \caption{{\it Left:} $J$ vs. $J-H$ diagram for luminous sources found within the IR-blobs and clumps  of the Sh2-90 complex. The ZAMS locus
reddened by \av $=$ 0, 10, 20, and 30 mag is shown in vertical dotted dashed lines.
Slanting  solid lines represent the standard reddening vector drawn from the ZAMS locus corresponding to
different spectral types. {\it Right:} $J-H$  vs. $H-K$ diagram for the same sources. The thin solid (blue) and
thick dashed (green) lines represent the unreddened MS and giant
branches (Bessell $\&$ Brett 1988), respectively.
The dotted line (blue) indicates the locus of intrinsic T-Tauri stars (Meyer et al. 1997). 
The parallel dashed lines are the reddening vectors drawn from the tip of the giant branch
(left reddening line) and from the base of the MS branch (right reddening line).
The sources with  IDs represent the same sources as  in Fig. 12, except source $\#21$ because its colours ($J-H$=5.6, $H-K$=3.6)
 fall well beyond the range shown in the plot.
}
  \label{fig13}
\end{figure*}
For four luminous embedded  YSOs (IDs 10, 13, 21, and 25), for which we have well-sampled
fluxes from NIR to 350 $\mu$m, we fitted models of Robitaille et al. (2007).
These models are computed using Monte Carlo based radiation transfer code
using several combinations of central star, disk, in-falling envelope, and bipolar cavity 
for a reasonably large parameter space. 
The fluxes of these four luminous sources at {\it Herschel} bands have been taken from the Curvature Threshold Extractor package (CuTeX)  catalog. 
The details about the catalog and the photometric procedures adopted in CuTeX  are 
given in Molinari et al. (2011, and references therein). 
While fitting the SED models, we adopted 10$\%$ to 30$\%$ errors in the flux estimates  and allowed distances 
in the range of 2.1-2.5 kpc. From SED models we constrained the key physical parameters such as 
stellar mass ($M_{\ast}$), stellar temperature ($T_{\ast}$),  disk mass ($M_{\rm disk}$), disk accretion rate 
($\dot{M}_{\rm disk}$), visual extinction ($A_V$), and the total luminosity ($L_{\rm bol}$).
The SED models of the four luminous  sources are shown in Fig. 14 and their physical parameters 
are tabulated in Table 6. The tabulated values are the weighted mean  and standard deviation of the parameters obtained from 
the best-fit models (i.e., models satisfying $\chi^2 - \chi^2_{\rm min} \leq 2N_{\rm data}$, where $\chi^2_{\rm min}$ is the
goodness-of-fit parameter for the best-fit model and $N_{\rm data}$ is the number of input observational
data points) weighted by e$^{({{-\chi}^2}/2)}$. The resolution of Herschel images is larger than the resolution of  NIR and {\it Spitzer} images (see Sect. 3),
thus the contribution from the surrounding environments to the flux estimates of these YSOs at {\it Herschel} bands (particularly $\lambda$ $\geq$ 160 $\mu$m), 
cannot be ignored. 
If this is the case, among the model-based parameters,  the parameter total luminosity is likely to be affected
most and is probably an overestimation. Similarly, inclination angle can also vary the SED shapes, and so the parameters. Thus, 
because of to several aforementioned limitations, the model-based parameters are 
only indicative of stellar and circumstellar properties of the underlying stellar source.Nevertheless, from Table. 6 
it can be inferred that for all the four sources the model based parameters seem 
to be constrained well, except  source $\#21$ for which the disk accretion rate is not  well constrained.
The limitations of the SED based   models, while inferring physical parameters have been discussed in the Robitaille et al. (2008; see also Deharveng et al. 2012). 
For example, for embedded Class 0/I sources,  
the Robitaille et al. models omit dust temperatures below 30 K, whereas the average envelope temperature of 
Class 0/I YSOs are generally lower (e.g., $\sim$ 15.7 $\pm$ 1.7 K in the W5 complex; Deharveng et al. 2012). In such cases, the models
overestimate the envelope mass ($M_{env}$)  of a YSO to account for fluxes at longer wavelengths. 
 The four luminous YSOs have been detected in the NIR and MIR bands, thus protostars have formed. 
The disk is a significant contributor to the SED shape at $\lambda$ $\le$ 100 $\mu$m (Whitney et al. 2005). We are mostly interested in the envelope mass.
We thus computed the $M_{env}$ of the luminous YSOs by fitting a modified black-body  to the  observed fluxes at 160, 250, 350, and 500 $\mu$m in order to 
avoid the  contribution of warm dust at  70 $\mu$m (due to internal heating from the  protostar and  emission from the disk).
The modified black-body  fits of the four luminous YSOs are shown 
in Fig. 15 and the derived $M_{env}$ values  are tabulated in Table 6. 

In the following, we  discuss the nature of the luminous sources 
within the clumps and IR-blobs. 
\begin{figure}
\centering{
\includegraphics[width=4.0cm] {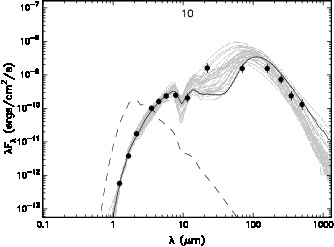}
\includegraphics[width=4.0cm] {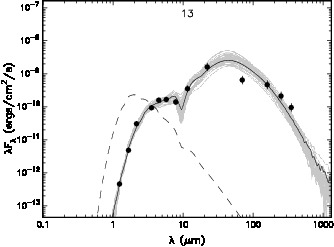}
\includegraphics[width=4.0cm] {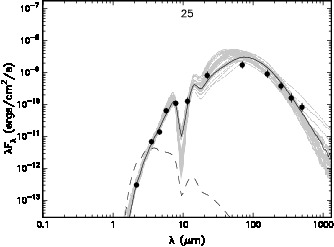}
\includegraphics[width=4.0cm] {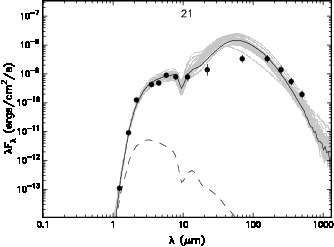}}
\caption{The Robitaille et al. (2007) SED models of the four  luminous embedded YSOs in the region (see text). The black line shows
the best fit, and the gray lines show subsequent good fits that satisfy $\chi^2 - \chi^2_{\rm min} \leq 
2N_{\rm data}$. 
 The adopted flux errors  are in the range of 10$\%$ to 30$\%$.
The
dashed line shows the stellar photosphere corresponding to the
central source of the best fitting model. 
The source of adopted flux values at NIR, GLIMPSE, and {\it Herschel} bands is discussed in the text.}
\label{fig14}
\end{figure}
\begin{figure}
\centering{
\includegraphics[width=4.0cm] {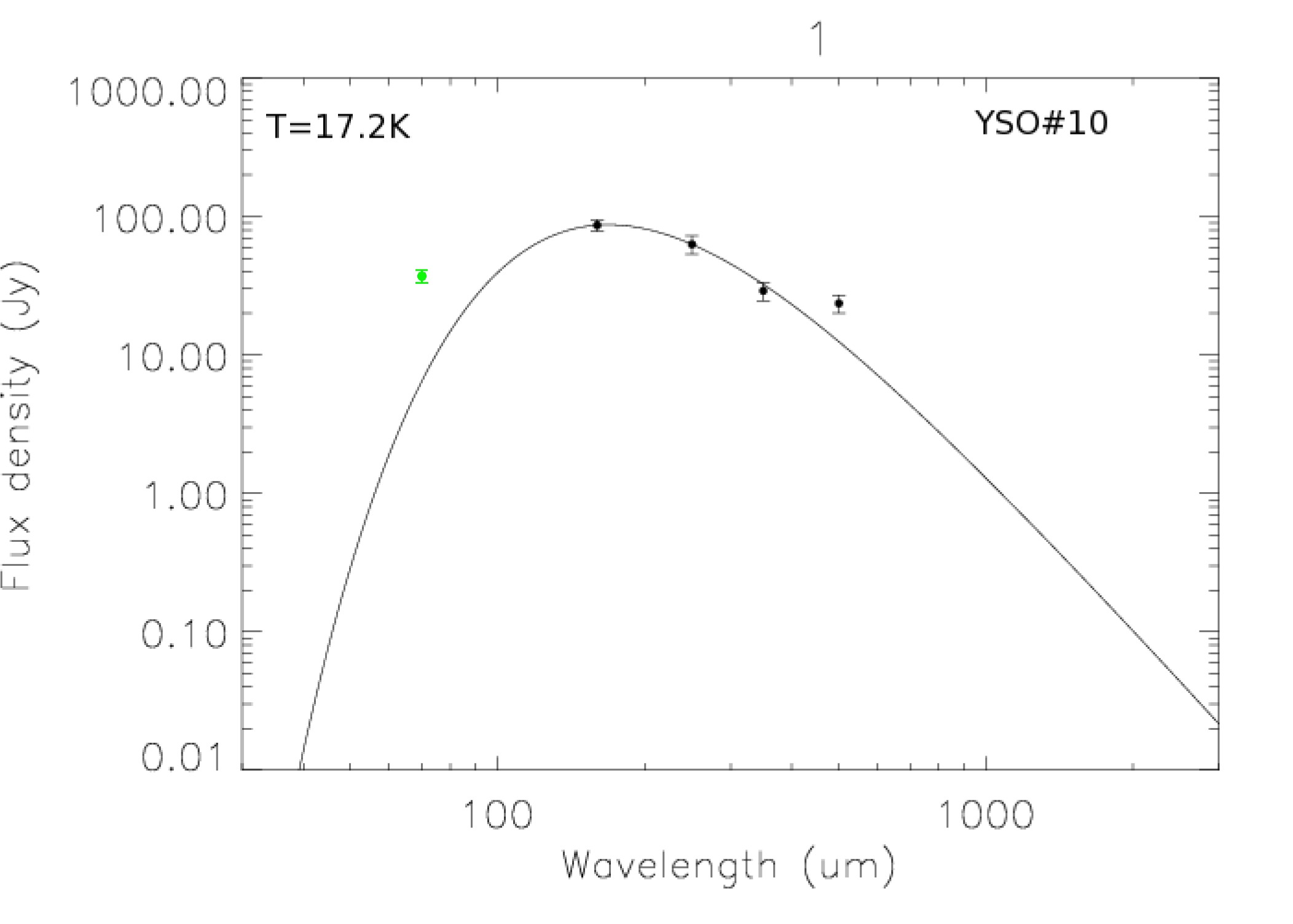}
\includegraphics[width=4.0cm] {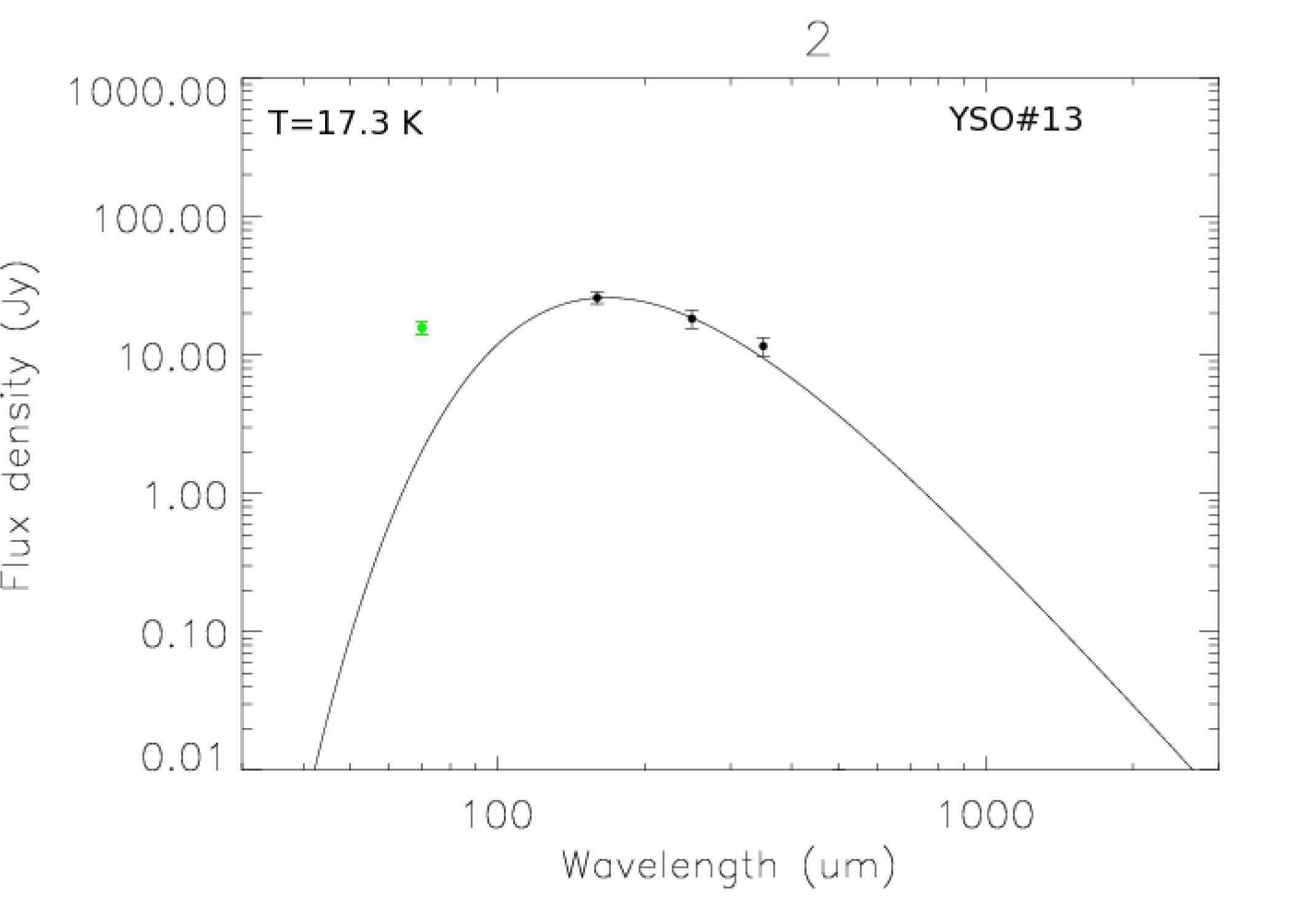}
\includegraphics[width=4.0cm] {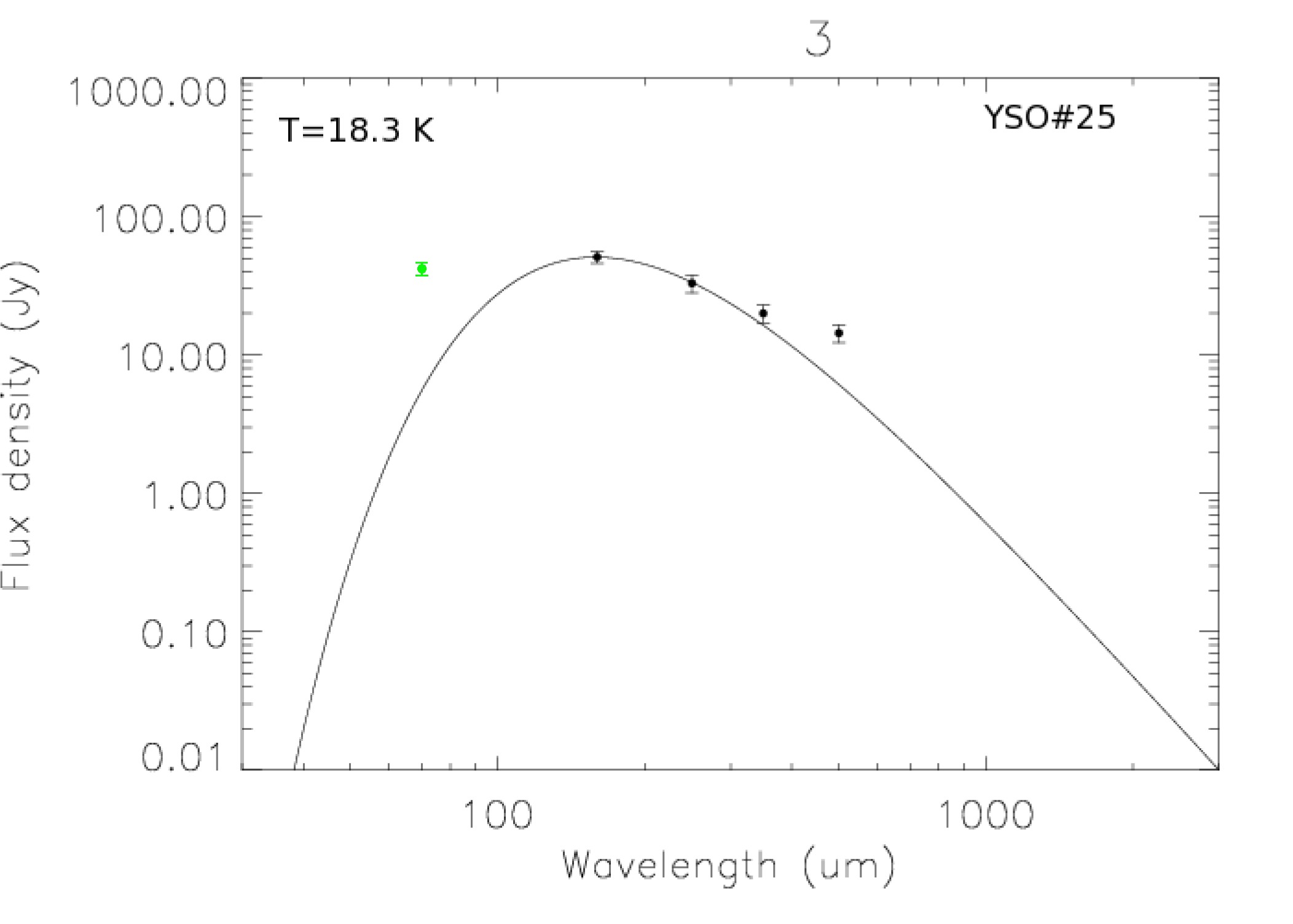}
\includegraphics[width=4.0cm] {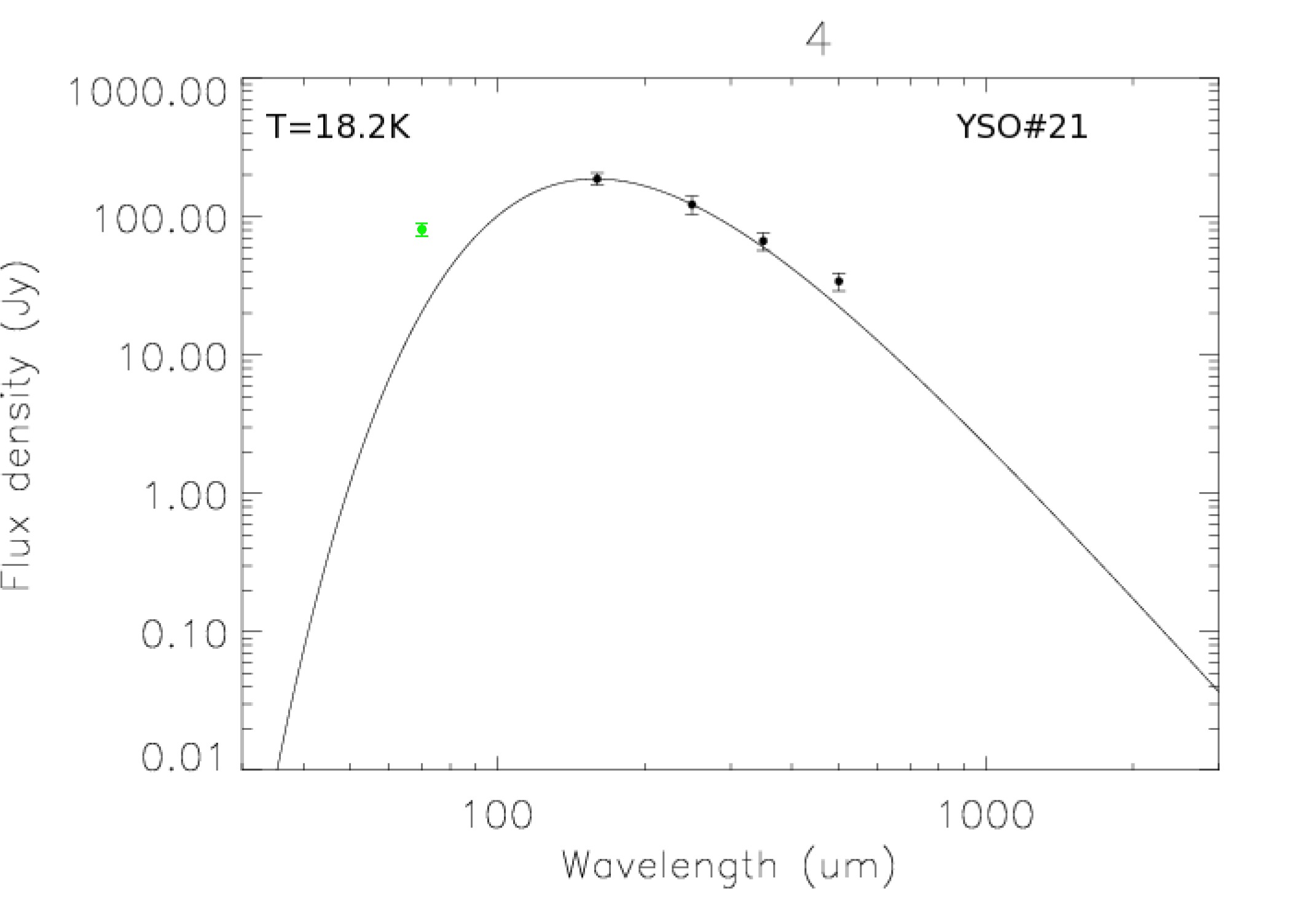}}
\caption{The grey-body SEDs of the envelopes of the four luminous embedded YSOs in the region. The black line shows
the best modified black body fit to the data points, between 160 $\mu$m and 500 $\mu$m. The circles denote the
input flux values.}
\label{fig15}
\end{figure}
\begin{figure}[htp]
\centering{
  \includegraphics[trim=0cm 0cm 0cm 0.0cm, clip=true, angle=0,width=8.5cm,height=11.0cm ]{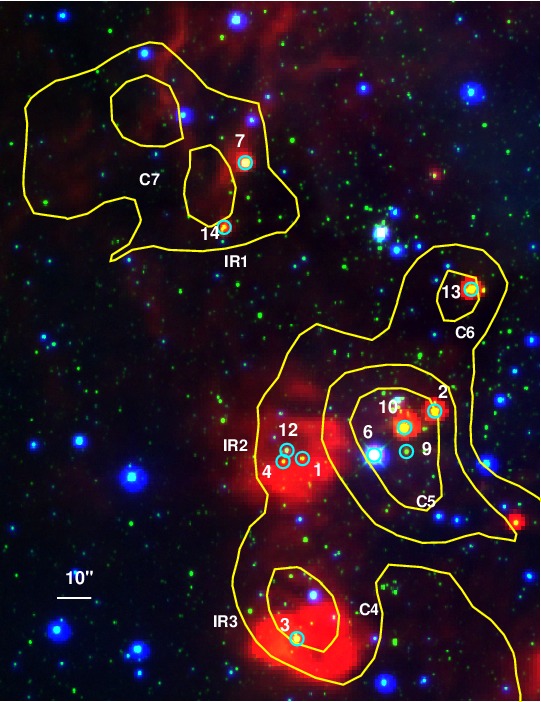}}
  \caption{ {\it Left:} The colour-composite image of the regions IR1, IR2, IR3, C5,  and C6 at 5.8 $\mu$m (red), 2.14 $\mu$m (green), 
and 0.80 $\mu$m (blue), with 250 $\mu$m contours. The contour  levels are at 1450, 2000 and 3000 MJy sr$^{-1}$. 
North is up and east is  left.}
\label{fig16}
\end{figure}

{\bf Clump C7}:  This is a 250 $\mu$m clump of mass $\sim$ 95 (89) \msun. In the western direction of the clump a  24 $\mu$m roughly circular 
(diameter $\sim$ 24$\arcs$) diffuse dust emission (see Fig. 2) is observed. 
The YSOs found in the direction of C7 are distributed along the ionization front (PDR seen in 8 $\mu$m and 70 $\mu$m) of Sh2-90. 
The luminous IR objects in the proximity of C7 are marked (labelled  $\#7$ and $\#14$) 
in Figs. 12 and 16.  Source $\#7$ is  luminous in all IRAC bands. Its position in the $J$ vs. $J-H$ 
diagram indicates an early B star, reddened by  \av $\sim$ 16 mag. 
 This source shows weak IR excess in the $J-H$ vs. $H-K$ diagram and no compact radio emission is detected in its direction, 
thus the possibility that it appears luminous because of excess emission at the $J$ and $H$ bands exists. Hence the present luminosity (mass) of the source is possibly an upper-limit. 
For a PMS star of  an assumed age of 1 Myr (see Sect. 6.1), its location on the HR diagram suggests  that it is a 
source of mass greater than 4\msun. Future spectroscopic observation would reveal the exact nature of the source.
 The position of source $\# 14$ in Fig. 13 suggest
that it is probably an intermediate-mass (B8-type) YSO with extinction $\sim$ 18 mag.
Clump C7  consists  of two 
sub-clumps. The two sub-clumps probably follow two PDRs seen in the 8 $\mu$m and 70 $\mu$m map.

{\bf Region IR2}: The region displays a circular (diameter $\sim$ 22$\arcs$) diffuse dust emission
in the wavelength range 5.8-70 $\mu$m. The luminous sources found within IR2 are
marked (labelled as  $\#1$,  $\#4$, and  $\#12$) in Figs. 12 and 16. 
None of them shows  sign of NIR excess. Two sources,  sources $\#1$ and $\#4$, appear to be reddened background objects,
as their position on $J-H$ vs. $H-K$ diagram falls close to the reddened giant locus.  
The position of  source $\#12$ on the $J$ vs. $J-H$ diagram (Fig. 13)  indicates a star of B7-type,
reddened by \av $\sim$ 10 mag. This source is the most-likely star responsible for the 
excitation of the 5.8-70 $\mum$ dust in IR2.
This region is devoid of cold dust emission at $\lambda$ $\geq$ 160 $\mu$m. 

{\bf Region IR3}: The morphology of IR3 is similar (diameter $\sim$ 20$\arcs$) to IR2 in the range 5.8-70 $\mu$m. In NIR, 
we detect a bright point source (labelled as   $\#3$ in Figs. 12 and 16) at the center of IR3.
The position of this source on NIR diagrams (Fig. 13) indicates a star of
spectral type close to B3, extincted by  \av $\sim$ 13 mag with no NIR excess. 
Being situated at the center of the nebula, it is the most-likely heating source of IR3. 
 A 250 $\mu$m  clump (C4) of mass 26(20) \msuns lies adjacent to IR3.

{\bf Clump C5}:  This is a 250 \mum  clump of mass $\sim$ 125 (75)
\msun. In NIR, we identified four bright sources (labelled as $\#2$, $\#6$, $\#9$, and $\#10$ in Figs. 12 and 16) 
close to the peak of C5, among which two sources  ($\#2$ and $\#10$)  show strong IR excess and 
appear luminous in $J$ vs. $J-H$ diagram, with spectral type earlier than B2 and extinction greater 
than 27 mag (if purely photospheric). Since both the sources show strong NIR excess, their stellar luminosity 
based on $J$ vs. $J-H$ is not reliable. Both the sources are of Class 0/I in nature; however, the emission at 
{\it Herschel} wavelengths, is mainly dominated by  source $\#10$.  The parameters (ID $\#10$ in Table 6) obtained from 
the best fit Robitaille et al. (2007) models (top left in Fig. 14) suggest it is a  $\sim$6 \msuns star 
of total luminosity of $\sim$0.7 $\times$ 10$^{3}$  \lsun~ embedded in a cloud of \av~$\sim$ 16 mag. 
We also fitted the SED models to  source $\#2$  with input fluxes in the range 1.25-22.0 $\mu$m to 
constrain some of its basic parameters. The models suggest it is a $\sim$ 5 \msuns star of total 
luminosity $\sim$0.9 $\times$ 10$^{2}$ \lsuns  embedded in a cloud of \av $\sim$ 18 mag.
Source $\#6$ is optically bright, highly luminous in $J$ vs. $J-H$ diagram with no excess, and is located close to
the unreddened giant locus in the $J-H$ vs. $H-K$ diagram, thus likely  to be a foreground field star.

Towards  C5, dense molecular tracers such as CS  (Beuther et al. 2002) and  $\rm {NH_3}$ (Wu et al. 2006) have been observed. 
The  signature of high velocity CO gas (Yang et al. 2002) has 
also been reported towards C5. We identified a significant number of YSOs
in the close proximity of C5 (see Fig. 11), thus possibly suggesting  a cluster forming site. \\

{\bf Clump C6}: This is a 250 \mum  clump of mass $\sim$ 31 (17) \msun~ located towards the projected center of the N133 bubble. The luminous IRAC source 
associated with C6  (labelled as $\#13$ in Figs. 12 and 16) is a Class 0/I YSO. Its position  
on NIR diagrams (Fig. 13), indicates a star of spectral type earlier than B1, extincted by \av $>$ 32 mag (if purely
photospheric). This source shows  strong NIR excess, thus its nature is uncertain. 
The parameters (ID $\#13$ in Table 6) obtained from the best-fit Robitaille et al. (2007) models (top right of Fig. 14) suggest 
it is a  $\sim$7 \msuns star with total luminosity of $\sim$0.8 $\times$ 10$^{3}$  \lsun, embedded in a 
cloud of \av~$\sim$ 12 mag.

\begin{figure}[htp]
\center
  \includegraphics[trim=0cm 0cm 0cm 0cm, clip=true, angle=0,width=9cm ]{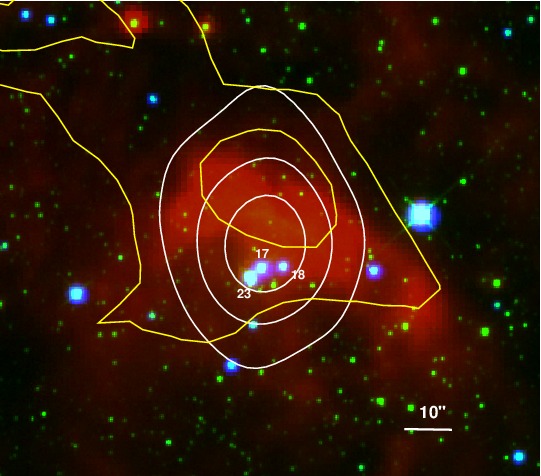}
  \caption{The colour-composite image of the region IR4 at 5.8 $\mu$m (red), 2.14 $\mu$m (green), and 0.80 $\mu$m (blue), overlaid 
  with 250 $\mu$m contours (yellow). The contour levels are at 1450  MJy sr$^{-1}$ and 2000  MJy sr$^{-1}$. 
  The white contours represent the radio  continuum emission at 1280 MHz. 
 North is up and east is left.}
  \label{fig17}
\end{figure}

{\bf Region IR4}: 
The region shows a cometary structure at 8.0 $\mu$m, partly protruding into the \hii region. 
The location of IR4 coincides with IRAS 19471+2641 of luminosity  $\sim$ 7.7 $\times$ 10$^3$ \lsun. 
In our high-resolution radio continuum map, we identified a compact 
radio source (shown in Fig. 17) of flux density $\sim$ 0.04 mJy, which corresponds to ionizing photon flux from a star of 
spectral type  B1 V (Smith et al. 2002). 
Three bright point sources (labelled as $\#17$, $\#18$, and $\#23$ in Figs 12 and 17) are found  inside
it. Among  these sources, the source $\#23$ is the most luminous source in NIR, and 
its location on NIR diagrams (Fig. 13) suggests that it  is a source of spectral type B2-B3, reddened by 
\av $\sim$  7 mag. The other two sources also  appear as B-type (B5-B7) stars.
This compact \hii region is most probably ionized by source $\#23$.
The hot dust appears as  circular and extended  at 24 $\mu$m and 70 $\mu$m around 
these B-type stars. A  PDR  is seen around these sources at 8.0 $\mu$m or 5.8 $\mu$m.  We identified a 250 \mum  
clump (C3) of mass $\sim$ 37 (28) \msuns adjacent to  IR4. The PDR is bright in the direction of C3.
\\

{\bf Clump C9}: This is a 250 $\mu$m  clump of mass $\sim$ 19 (16)
\msuns located on the PDR of Sh2-90. This region coincides with the position of  IRAS 19470+2643 of 
luminosity $\sim$ 7.6 $\times$ 10$^3$ 
$\lsun$. The luminous sources near C9 are marked in Fig. 12 (labelled as $\#5$, $\#11$, and $\#24$).
Source $\#11$ appears to be a B5 in the $J$ vs. $J-H$ diagram; however its location is
close to the unreddened giant locus in the $J-H$ vs. $H-K$ diagram, thus it is probably a foreground field star.
The position of the sources $\#5$ and $\#24$ on the $J$ vs. $J-H$ diagram (Fig. 13) shows the characteristics of a spectral 
type earlier than B3, with \av $>$ 19 mag. These sources do not show NIR excess, and  the maximum column density 
in the direction of C9 suggests an extinction of $\sim$ 9 mag. These sources are slightly relatively more reddened 
than the other massive sources associated with the regions such as IR1, IR2, IR4, and IR4 observed in the PDR of Sh2-90. 
No 8 $\mu$m or 24 $\mu$m symmetric emission structures have been noticed around C9. Thus, we suspect 
that the bright NIR sources of C9 are probably reddened background objects, which is also supported by the location
of $\#24$ on the $J-H$ vs. $H-K$ diagram.

\begin{figure}[htp]
\centering{
  \includegraphics[trim=0cm 0cm 0cm 0.0cm, clip=true, angle=0,width=8.5cm,height=8.5cm ]{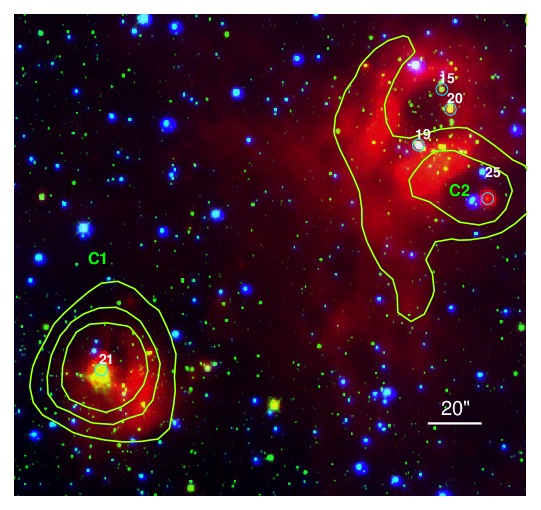}}
  \caption{Colour-composite image of the regions N132 and C1 at 5.8 $\mu$m (red), 2.14 $\mu$m (green), and 0.80 $\mu$m (blue), 
  overlaid with 250 $\mu$m contours. North is up and east is  left.}
\label{fig18}
\end{figure}

\begin{figure}[htp]
\centering{
  \includegraphics[trim=0cm 0cm 0cm 0.0cm, clip=true, angle=0,width=8.6cm,height=5.5cm ]{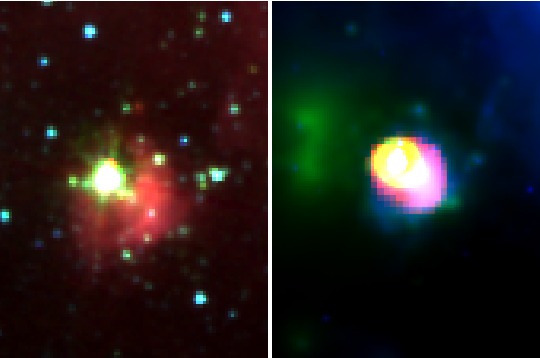}}
  \caption{ {\it Left:} Colour-composite image of the region C1 at 5.8 $\mu$m (red), 4.5 $\mu$m (green), and
 3.6 $\mu$m (blue). {\it Right:}  Colour-composite image of the same region  at 70 $\mu$m (red), 24 $\mu$m (green), and 8.0 $\mu$m (blue).
North is up and east is  left.}
\label{fig19}
\end{figure}

{\bf Region N132/C2}: 
N132 is roughly a circular  bubble of diameter $\sim$ 0$\farcm$28, with a central cavity and a thin annular shell bordering it. 
The shell is best seen at 8.0 $\mu$m.  Unlike N133, it is not visible in the optical.
No 23 cm emission has been observed towards N132, thus 
this infrared dust bubble was possibly created by a less massive star than the O-B2 stars.  The luminous sources 
(labelled as $\#15$, $\#19$, and $\#20$) 
projected inside the bubble are marked in Figs. 12 and 18. The position of  source 
$\#19$ on the $J$ vs. $J-H$ diagram suggests 
it is a star of spectral type close to B3, extincted by $\sim$ 9 mag, whereas the 
location of other sources ($\#15$ and $\#20$ in Fig. 18) possibly suggests that they are  cooler stars; because these stars 
possess IR-excess, they appear luminous in the $J$ vs. $J-H$ diagram. 
A  clump (C2) of mass $\sim$ 43 (37)\msuns lies 
close to the south-western boundary of N132 (see Fig. 18). A source ($\#$ 25) at the center of this clump is visible 
in all the IRAC bands, and its IRAC colours are consistent with a Class 0/I YSO. This source is not detected 
in $J$ and $H$ bands, thus is highly embedded.  The parameters (ID $\#25$ in Table 6) from the best-fit SED models 
(bottom left in Fig. 14) suggest it is a $\sim$6 \msun~ star with a total luminosity  $\sim$0.6 $\times$ 10$^{3}$  \lsun~ 
embedded in a cloud of \av~$\sim$ 42 mag.

{\bf Clump C1}:  This is a clump of mass 206 (186) \msun. The clump is bright at {\it Herschel} bands. It is associated with a point source (ID $\#21$ in Fig. 18)
seen in all wavelengths between 1.22 $\mu$m and  500 $\mu$m. At longer  wavelengths ($\geq$ 250 $\mu$m) 
elongated extended emission (i.e., in the direction of f1; see Fig. 3) is seen in the N-W direction of C1. A faint diffuse 
24 $\mu$m emission can also be seen in its N-E direction (see Fig. 19). This source shows strongest  IR-excess in the $J-H$ vs. $H-K$ (i.e., $J-H$ = 5.6 and $H-K$ = 3.6) 
CC plot and is the most luminous one in the $J$ vs. $J-H$ diagram  (it does not appear in Fig. 13). Its IRAC colours suggest that it is a 
Class 0/I YSO. The parameters (ID $\#21$ in Table 6) obtained from the best-fit Robitaille et al. (2007) models (bottom right in Fig. 14)
suggest it is a  $\sim$ 8.8 \msuns star with a total luminosity of $\sim$4.0 $\times$ 10$^{3}$  \lsun~ embedded in a cloud  
of \av~$\sim$ 31 mag.

The location of C1 coincides with IRAS 19747+2637 of luminosity $\sim$1.4 $\times$ 10$^{4}$ \lsun.
No 1.3 cm continuum emission is detected towards IRAS 19747+2637 at rms of 0.7 mJy beam$^{-1}$ (Wang et al. 2007).
The 0.7 mJy flux  corresponds to the number of Lyc photons of $\sim$ 2.1 $\times$ 10$^{44}$ $s^{-1}$ (spectral type later than B2 V). 
Therefore, the present upper limit on mass of the massive source  expected to be $\sim$ 10 \msun. 
The source IRAS 19747+2637 is associated with a \wat maser (Codella \& Felli 1995). Observation in the CO line (J=1-0,  Xu et al. (2001); 
Yang et al. 2002) suggested that IRAS 19747+2637 is associated with a high-velocity gas, thus is an outflow candidate; SiO emission has been detected in this region, which increases the possibility of shocks associated with massive 
young sources (Harju et al. 1998). 
The detection of a maser and signature of shock suggest that the luminous YSO is at its early evolutionary phase. The cloud containing this 
luminous YSO also harbors many low-mass YSOs (with masses $\le$ 3 \msun). The detection of a
large number of YSO candidates in the close vicinity of C1 suggests  a site of cluster mode of massive star 
formation, where the massive star has not yet  developed an ultra compact \hii region. This is probably a site of  
young clusters in the process of formation.

\begin{figure}
\includegraphics[width=6.3cm,height=8.5cm,angle=270 ]{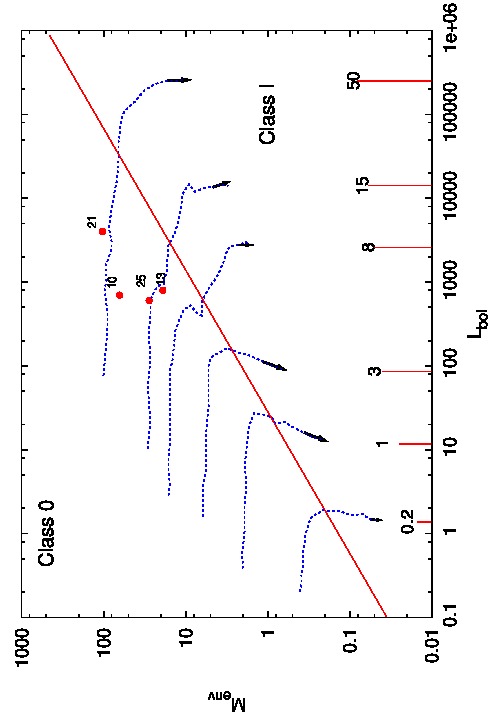}
  \caption{L$_{\rm bol}$ vs.  M$_{\rm env}$ diagram for the massive YSOs (solid circles with IDs).
      The dotted blue lines represent the evolutionary tracks from Andr\'e et al. (2008).  Evolution proceeds from the upper left to the lower right (indicated by arrows at the end of each track). The final stellar masses of these tracks in solar units 
       are given above the lower axis. The slanted red  line corresponds to the location where 50\% of the 
      initial core mass being converted into stellar mass (c.f.  Bontemps et al. 1996;  Andr\'e et al. 2000).
      }
\label{fig20}
\end{figure}

\subsection {Evolutionary status of massive sources}

The SED models suggest that the IR sources (IDs $\#10, \#13, \#21$, and $\#25$) are massive (6-9 \msun)  and of high luminosity (0.4-4 $\times$ 10$^{3}$ \lsun). 
These sources are classified as Class 0/I YSOs based 
on IRAC data; however, there are scopes to revise  further the classification of individual Class 0/I YSOs 
in the light of {\it Herschel} observations.
For these sources, we found high values of sub-millimeter to bolometric luminosity 
(L$_{\lambda \geq 350 \mu{\rm m}}$/L$_{\rm bol}$ $>$ 0.2 \msun/\lsun), envelope mass to the mass of central 
protostar (M$_{env}$/M$_{\star}$ $>$ 3.4), and envelope mass to bolometric luminosity (M$_{\rm env}$/L$_{\rm bol}$ $>$ 0.03), 
indicating they are more likely to be Class~0 protostars (see Andr\'e 2000) in their accretion phase. 
We also use the physical parameters obtained for the sources to infer their evolutionary status 
using M$_{\rm env}$--L$_{\rm bol}$ diagrams (Andr\'e et al. 2000; Molinari et al. 2008).
The location of these four sources on the M$_{\rm env}$--L$_{\rm bol}$ evolutionary 
diagram for protostars is shown in Fig. 20, and their comparison with the evolutionary tracks of 
Andr\'e et al. (2008 and references therein), suggests that these objects will evolve into stars of mass $\geq$ 15 \msun. 
In summary, these YSOs are massive YSOs (MYSOs) of the region likely to be in their Class 0 or early Class I stage. 
  It is especially evident for $\#10$ and $\#21$, as 
signatures of outflow (e.g., high velocity CO gas or SiO emission) and/or maser emission have been observed in both  cases (Xu et al. 2001; Harju et al. 1998; Yang et al. 2002).

Are these four MYSOs  in  different evolutionary stages than the massive sources 
associated with the IR-blobs (IR1, IR2, IR3, and IR4)? In the absence of the 
stellar properties (e.g., stellar luminosity and temperature)  and 
the lack of sensitive longer wavelength observations to model the SEDs of the massive sources in the IR-blobs, a clear distinction 
cannot be made. However, from the available physical conditions of their environment and stellar properties, the evolutionary 
status of these two groups can be inferred up to a certain extent.
It is expected that the mass of the cold components decreases with time as the protostellar envelope accretes onto
the central object and progressively the central massive protostar emerges from the cloud by dispersing its natal
environment.  Although the four MYSOs are 
visible in NIR, they show strong NIR excess and a large fraction of their luminosity lies in the far-infrared to sub-millimeter regime (see Fig. 14).
They are surrounded by cold components and  thus possibly actively accumulating material from the cold envelope. The massive source(s) of IR-blobs 
are optically visible, show little or no IR excess, and cold components are often not found at their exact 
locations; all of which indicates that they have little or no circumstellar dust left. 
The above characteristics strongly suggest that the massive sources of the IR-blobs should represent an older evolutionary sequence than the 
envelope-dominated likely Class 0 MYSOs of the complex.
 
\section{Kinematics of ionized and molecular material, and  evolutionary status of  Sh2-90}
\begin{figure}[htp]
\centering
  \includegraphics[trim=0cm 0cm 0cm 0cm, clip=true, angle=0,width=7.5cm ]{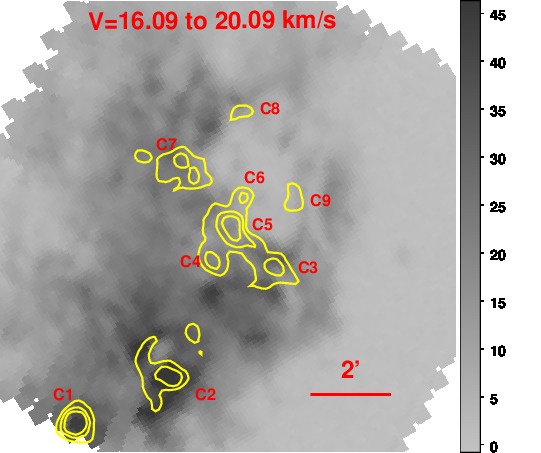}
  \includegraphics[trim=0cm 0cm 0cm 0cm, clip=true, angle=0,width=7.5cm ]{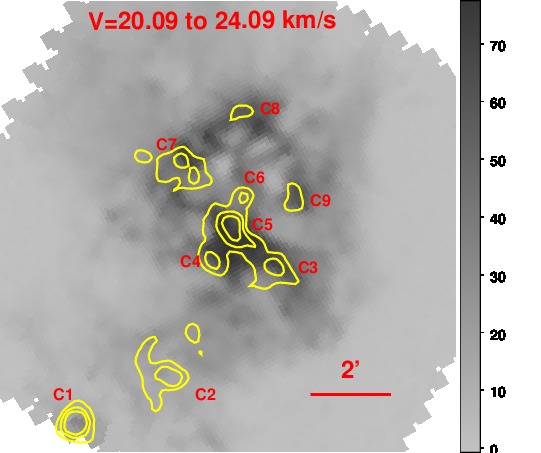}
  \includegraphics[trim=0cm 0cm 0cm 0cm, clip=true, angle=0,width=7.5cm ]{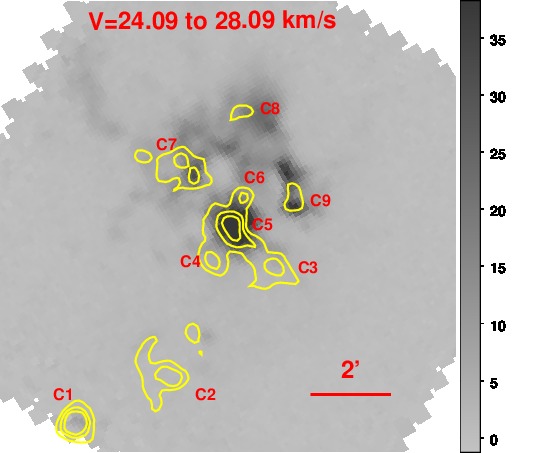}
  \caption{ The $^{12}$CO(J=3-2) integrated emission in the direction of Sh2-90 centered at
 $\alpha_{2000} = 19^{\rm h}49^{\rm m}14^{\rm s}$,
$\delta_{2000} = +26^{\circ}50^{\prime}01^{\prime\prime}$.  Velocity ranges are indicated at the top of each image.  The horizontal colour bars are  labelled in units of  K \kms.
  The data is  from  Beaumont \& Williams (2010; angular resolution of 16\arcs). The overlaid 250 $\mu$m contours are 
  at 1450, 2000, and 3000 MJy sr$^{-1}$. The discussed clumps are also marked (see also Fig. 12).  North is up and east is left.}
  \label{fig21}
\end{figure}
We used the $^{12}$CO (J = 3-2) data cubes of Beaumont \& Williams (2010) to study the association of the cold dust  
clumps discussed in Sect .5. The inspection of data cubes showed that the CO emission  over the whole field is mainly in the velocity
range 15.5 - 28.5 \kms~(only weak CO emission is seen outside this velocity range). We isolate the CO emission of the Sh2-90 complex along 
the line of sight mainly into three major components: i) 16.1 -  20.1 \kms~ ii) 20.1  - 24.1 \kms, and iii)  24.1  - 28.1 \kms. Figure 21 shows the 
distribution of integrated CO emission as a function of velocity in these velocity ranges:

$\bullet$ The 16.1 to 20.1 \kms~ component shows a  cavity corresponding to the location of Sh2-90 and 
an extended diffuse cloud at the east and south of Sh2-90. We also observed similar extended dust emission towards the east and south of Sh2-90
in our column density map (see Fig. 7). This velocity component  also displays prominent emission at the location of N132 (or C2) and IRAS 19474+2637 (or C1), 
thus confirming the association of these clumps with the complex.

$\bullet$ Gas in the interval from 20.1 \kms to 24.1 \kms~ shows enhanced molecular emission in a ring-like structure  that 
resembles the {\it Herschel} 250 $\mu$m dust shell observed at the periphery of Sh2-90 (see Fig. 8). This possibly represents 
the collected matter around the \hii region. The  clumps C3, C4, C7, C8, and C9 are projected on this ring-like 
CO emission, thus they are likely to be associated with Sh2-90. 
The feature south of C5  is more  prominent in the velocity interval 
from 21.1 \kms to  22.7 \kms~and for more positive velocities its peak emission shifts towards C5. 

$\bullet$ In the  24.1 to 28.1 \kms~ range, the  features of C7 and C9 remain visible; however, the strongest molecular 
emission at these velocities is mainly coincident with region C5.

$\bullet$ Clump C6 is projected towards the center of  Sh2-90; its association is difficult to determine because 
we did not notice any strong CO emission at the exact location of C6.

Lafon et al. (1983)  observed ionized gas emission towards Sh2-90, using the optical ${\rm H\alpha}$ line, in the velocity 
range \vlsr $\sim$ 8 - 33 \kms~ with a mean velocity of 20.8 \kms. 
Although their velocity map shows complex velocity distribution, which could be  due to the effect of strong irregular 
extinction across the region, the velocity distribution mainly shows \vlsr $>$ 26 \kms~  in the N-W side 
(corresponding to the brightest zone of optical emission) and   \vlsr  $<$ 14 \kms~ in the S-E side of Sh2-90. This velocity range 
is consistent with the velocity range of the molecular gas discussed above. The large velocity range covered by the ionized gas may be indicative of a ``champagne flow'' 
(Tenorio-Tagle 1979). Although in our low-resolution 610 MHz radio map, we see a weak gradient in brightness distribution 
from N-W to S-E direction, on a large scale the shape of ionized gas appears to be more symmetrical (see Fig. 3) with 
ionizing star(s) at the center of nebula. Thus the \hii region is possibly in its early  phase of~ ``champagne-flow''. 

 During the evolution of the \hii region, if the ionized gas expands spherically, then the 
difference in the extreme radial velocities (either blue shifted or red shifted) should roughly 
reflect its  expansion velocity.  If we consider the  velocity measurements at the two projected opposite edges of 
Sh2-90 (i.e., 26 \kms~  N-W and 14 \kms~ S-E) as the extreme 
radial velocity values of the \hii region, then this would indicate a crude expansion velocity of $\sim$ 6 \kms. 

Here,  we tried to derive the dynamical age of the \hii region, based on the observed properties and assuming 
that the \hii region evolved in a homogeneous medium.
The ionizing photon flux coming from the massive stars associated with Sh2-90 is  
$\sim$ 22.5 $\times$ 10$^{47}$ s$^{-1}$. The radius of the \hii region is $\sim$ 1.6 pc.
The density of the original medium is unobservable; only an approximate value can be adopted based on the
present observable indicators.  The average density of clumps in the shell 
is $\sim$ 7(5) $\times$ 10$^{3}$ 
cm$^{-3}$ (see Table 4), which possibly represents an upper limit (assuming it has purely formed from the collected matter). 
A lower limit on density of the original medium estimated
from the total ionized gas of the \hii region and neutral matter collected in the shell
is $\sim$ 2.6 (2.0) $\times$ 10$^{3}$ cm$^{-3}$ (see Sect. 5.3). 
Considering that the \hii region is expanding  at 
velocity $\sim$ 6 \kms, in a homogeneous medium  of density in the range  2-7 $\times$ 10$^{3}$ cm$^{-3}$, and  following 
the model by Dyson \& Williams (1997), we estimated the dynamical age (i.e., the time it would take to form 
from its initial Str\"omgren  sphere to its present size) in the range $\sim$ 0.6-1.2 $\times$ 10$^{6}$ yr.
We caution that this age estimations should not be taken too literally, 
because of the possible existence of a ``champagne-flow" and the assumption of evolution in a homogeneous medium, 
which is probably far from the truth. 
The MS lifetime of an O8 star is of the order 
of $\sim$ 6.5 $\times$ $10^6$ yr. This allows us to constrain the dynamical age  younger than 6.5 $\times$ $10^6$ yr. 

\section{Star formation scenarios in the complex}
Based on multi-wavelength observations presented in this work our understanding of star formation 
associated with the Sh2-90 complex can be summarized as follows:

$\bullet$ The Sh2-90 \hii region was created by an O8-O9 star (see Sect. 4.4). The O-type star possibly formed  1 Myr ago (see Sect. 7), created  
ionized gas around itself which then expanded, sweeping ambient ISM gas and dust into a bubble. The morphological correlation between ionized gas, PDR, and 
ring-like dust emission (see Sects. 4.2 and 5.3) around the ionized gas led us to believe that there is certainly a high degree of interaction between
the UV photons from the massive star and  the surrounding molecular cloud. 

$\bullet$  {\it Herschel} cold dust properties and {\it Spitzer} point source analyses reveal that star formation in the 
complex is not confined to the Sh2-90 \hii region (see Sects. 5.3 and 6.3).  Several sites of star formation have been observed in this
massive (M $>$ 10$^{4}$ \msun) cloud. The main star-forming sites are associated with the complex as they have similar velocities (see Sect. 8.5).
Based on the  velocities of the molecular and ionized gas, and signs of interaction of the Sh2-90 bubble with the molecular gas, we believe that
most of the extended cold dust emission seen at {\it Herschel} wavelengths is associated with the complex.

$\bullet$ The column density map presents an elongated structure similar to that seen on the low resolution (4\farcm4) 
\tco  map (Lafon et al. 1983).  We identified 129 likely YSO candidates, 
which includes 21 Class I, 34 Class II, and 74 NIR-excess YSOs. (see Sect. 6.1). The spatial distribution of YSOs 
follows the distribution of the high column density matter (see Sect. 6.3). As these young sources  have not had time to  
move away very far from their birthplaces, this  indicates that the elongation and sub-clustering of young sources 
 possibly resulted from  the primordial distribution of the parental dense gas in this complex. 

$\bullet$  We identified four Class 0/I MYSOs  in their main accretion phase (see Sect. 6.5). The simultaneous presence of envelope-dominated  Class 0 MYSOs (typical age $\sim$ few 10$^4$ yr; Maury et al. 2011) at various locations and  
of the ionizing star of an evolved \hii region (age $\geq$ 1 Myr; see Sect. 7) in the same complex suggests that different episodes of 
star formation have occurred in this complex. How star formation occurred at different locations 
in this extended cloud is difficult to determine. 
Large-scale SFRs often contain smaller size cores/clumps of high column densities possibly due to density fluctuations present in 
the original cloud. 
Star formation 
can occur in these cores/clumps on a local dynamical time scale as long as the local conditions are close 
to gravitationally bound and so  can result in  multiple stellar groups in the same complex, possibly in different  evolutionary stages.
However, since Sh2-90 is an evolved \hii region, we can also speculate that the formation of 
young sources in the region  might have been triggered by the expanding bubble. We will discuss this aspect 
in the following. However, smoothed particle hydrodynamic  
simulations (Dale et al. 2011, 2012) have shown that several untriggered stars can be formed in the extended environment of bubbles/\hii regions by other processes, including spontaneous ones.
 
Similar to the findings of Deharveng et al. (2010), Thompson et al. (2012; based on the Red MSX Source survey of MYSOs) suggested that the formation of only about 14\% to 30\%  of the massive 
stars in our Milky Way could have been triggered or influenced by \hii regions, whereas the remaining 86\% to 70\% stars are 
likely to be formed by other processes. 
Although most of the identified YSOs in the Sh2-90 complex are of low mass ($<$ 3 \msun; see Sect. 6.2),  67$\%$ of 
them are situated beyond the bubble projected radius. The Sh2-90 complex is  extended and clumpy (see Fig. 7, $bottom$), 
thus some of the YSOs could be the  result of spontaneous star formation. In particular, if we considered SFRs such as 
IRAS 19474+2637 (or C1) and IRAS 19473+2638 (or C2), then their locations beyond the main interaction zone of the 
bubble suggest that the star formation in these regions  is unlikely to be triggered by the expansion of Sh2-90.
We calculated that the
ionizing photon flux from  the ionizing star at the location of 
IRAS 19474+2637 ($\sim$ 4.7 pc away) and IRAS 19473+2638 ($\sim$ 3.2 pc away) is  less than  10$^{9}$ cm$^{-2}$ s$^{-1}$, 
the flux  required to trigger star formation even in a 
low-mass ($<$ 5 \msun) cloud (Bisbas et al. 2011). From the above evidence, we hypothesize that some of the 
YSOs (particularly the distant regions forming MYSOs) in the complex, 
could have formed spontaneously or by some other processes.

\begin{figure*}[htp]
\centering {
  \includegraphics[width=13.0 cm]{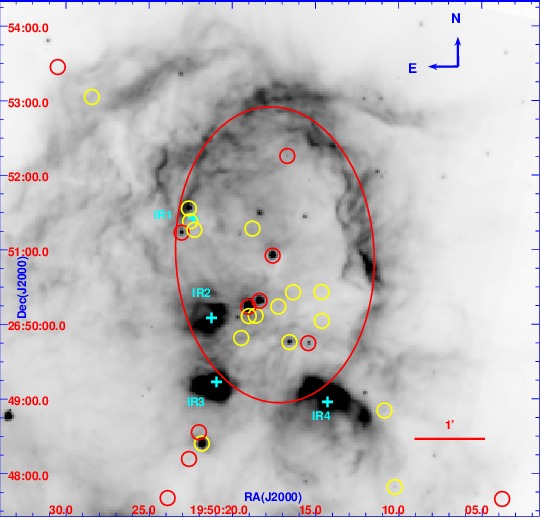}}
  \caption{
  The morphology of the region at 8.0 $\mu$m, over plotted with Class 0/I (red circles) and Class II (yellow circles) YSOs. 
  The position of the B-type sources are marked with  plus signs. The ellipse represents the projected boundary of 
  the 8.0 $\mu$m cavity.  
}
  \label{fig22}
\end{figure*}

$\bullet$ The star formation activity
at the periphery of Sh2-90 is somewhat different. We observed  that young B-type stars  associated with 8-24 $\mu$m compact circular structures  such as  IR2, IR3, and IR4   
lie at the  edge of Sh2-90. Our search in a wider area with  IRAC and MIPS images, resulted in no such compact, 
spatially closed circular structures, thus their chance of alignment along the line of sight is unlikely. 
They are possibly young (supported by the compact nature of their radio and/or 8-24 $\mu$m  dust emissions), although this
cannot be warranted without spectroscopic observations.
There are also seven distinct clumps C3, C4, C5, C6, C7, and C9 (the clumps C3, C4, and C7 are 
located adjacent to  IR4, IR3, and IR1, respectively). The clumps C3, C4, C7, and C9 are part of the connecting shell of cold material which 
partially borders the 8 $\mu$m cavity.  The distribution of B-type stars associated with IR structures and/or cold clumps around the  periphery of 
Sh2-90 (age $\sim$ 1 Myr), thus  
suggests that star formation has been possibly triggered in such regions. The case of C5 and C6 with MYSOs is not so clear.
These clumps are seen in the direction of the ionized gas; they probably lie in front of it, as they are seen in absorption 
at optical bands. Thus the formation of the YSOs they contain can also be triggered. However, the velocity of C5 is more positive 
with respect to the  mean velocity of the Sh2-90, and the evolutionary status of the  clump C5 is  similar 
to C1 and distinctly different from the sources within the IR-blobs (see Sect. 6.4 and 6.5). Thus, the possibility 
that this could also be one  clump 
like C1 located outside the action zone of Sh2-90 and going through star formation by some other processes cannot be excluded.

Proving the nature of triggered star formation around  bubbles/\hii regions formed in a non-uniform extended cloud is difficult,  
as   triggered sources can be  mixed  with the spontaneously forming ones. The complex environment of 
Sh2-90 is a subject of concern, nonetheless, on a large scale the distribution of neutral matter in a  shell-like structure, 
with massive clumps at regular intervals at its edges, suggests that the triggering might have happened through the  
collect and collapse (Elmegreen \& Lada 1977; Deharveng et al. 2005; Zavagno et al. 2006) process of star formation, in which
fragmentation occurs due to  large-scale gravitational instabilities of the material accumulated during the 
expansion of the \hii region. 
We explore the 
feasibility of the collect and collapse process by comparing the observed properties with the predictions of the analytical 
models by Whitworth et al. (1994).

  To compare, we adopt $N_{Lyc}$ $\sim$ 22.5 $\times$  10$^{47}$ s$^{-1}$ from radio continuum observations
and $\sim$ 2(7) $\times$ 10$^3$ cm$^{-3}$  as the density of the original medium (see Sect. 7) inside which the \hii region
evolves.
We consider the  sound speed in the collected layer as $\sim$ 0.3 \kms, corresponding 
to an average temperature of the cold dust observed in the shell (see Sect. 5.2). 
With these values, the model predicts the fragmentation of the collected material after $\sim$  1.6(1.0) Myr 
at a radius of  $\sim$  4.0 (2.0) pc, with 
fragments separated by some  $\sim$ 1.3(0.8) pc and having mass   $\simeq$ 84(47) \msun. 
These predicted values are comparable with  the observed properties, such as the dynamical age ($\sim$ 1 Myr) 
of the \hii region, the  separation ($\sim$ 0.9 pc) between the B-type stars and the bound fragments, and  the mass of the 
bound fragments ($\sim$ 17-125 \msun)  at the periphery of the bubble. However, the projected radius ($\sim$ 1.3 pc) at
which B-type stars 
have been observed is smaller than the minimum radius $\sim$ 2.0 pc predicted by the model. The above comparisons were done by simplifying facts throughout, 
as the caveats such as champagne flow, non-homogeneous medium, and projection effects always play crucial roles in such comparisons. 
Taking the present observational evidence at face value, the possibility of the young sources observed at the periphery
of Sh2-90 might have formed through the collect and collapse process exists.

 Alternatively, the Sh2-90 complex could be a case of star formation  as found in the simulations of Dale et al. (2012) for bound turbulent clouds 
under the influence of internal ionizing  sources.
In their simulations Dale et al. (2012) infer objects  which are triggered in a cloud complex 
by comparing the cloud with  feedback sources with a control cloud without feedback sources. 
In all their simulations the triggered stars are found to be embedded in dense gas on the borders of feedback
driven bubbles, and at the tips of pillars; but they are found to be mixed with the spontaneously 
formed ones. In the Sh2-90 complex   B-type stars are formed at the periphery of the \hii region (see Fig. 22), but many 
low-mass ($\leq$ 3 \msun) sources of similar evolutionary nature are found to be formed far away  as well as in the 
close vicinity of the \hii region (see Sect. 6.3).
As discussed above, star formation seen at far away locations could be spontaneous (particularly at the location of C1).
Thus, it is difficult to demonstrate that the formation of a specific source close to the \hii region is triggered by 
the \hii region rather than spontaneously;  the geometrical distribution of some of the  YSOs and B-type stars close to the 
rim of the bubble (see Fig. 22), association with the cold gas (see Sects. 6.4 and 7) and their youthness; however, suggest that their
formation is possibly triggered by the \hii region.

To put the star formation scenario of the complex on a firm footing, detailed velocity and age measurements 
of the sources in the complex are needed.

\section{Conclusions\label{con} }
In this paper we have presented {\it Herschel} and radio continuum imaging  of the  Sh2-90  complex as well as
{\it Spitzer}-IRAC and deep NIR imaging to explore its stellar and interstellar content, and star formation history. Based on these observations our results can be summarized as follows:

1. The Sh2-90 complex consists of two bubbles (N133 and N132) and a few IRAS  sources at various locations. N133 is a large bubble 
outlined by {\it Spitzer} 8.0 $\mu$m emission and together with Sh2-90, encloses the main \hii region of the complex.
It is an evolved \hii region of diameter $\sim$ 3.2 pc with an rms electron density $\sim$ 144 cm$^{-3}$, and an ionized mass $\sim$ 55 \msun. 
Our NIR photometry of the sources inside the bubble reveals the presence of a loose cluster, with the most massive 
member of which is a O8-O9 V star that is responsible for the ionization of N133.

2. The column density and temperature maps constructed from the {\it Herschel} observations suggest that Sh2-90
is  part  of a massive ($\ge$  10$^{4}$ \msun) elongated cloud of column density $\geq$ 3 $\times$ 10$^{21}$ cm$^{-2}$.
We  observed that neutral collected material of mass $\ge$ 637 $\msun$
is present in a shell surrounding the \hii region. Nine clumps
are detected in the complex, among which seven (mass range 8-125 $\msun$) are located at the periphery or in the direction of  Sh2-90. 
Four of them are co-spatial with  B-type stars and a compact \hii region. The velocity information of the clumps derived from CO (J=3-2)
data cubes suggests that most of them are likely to be associated with the Sh2-90 complex.

3. Using the IRAC and NIR CC diagrams, we identified 129 likely YSO candidates of  masses in the range of 0.2-3 $\msun$, 
which includes 21 Class I, 34 Class II, and 74 NIR-excess YSOs. 
 The photometric  measurements of these YSOs are available in electronic form at the CDS
via anonymous ftp to cdsarc.u-strasbg.fr (130.79.128.5) or via
http://cdsweb.u-strasbg.fr/cgi-bin/gcat?J/A+A.
We identified four Class 0 MYSO  in their main accretion phase.  
We observed that the spatial distribution of the candidate YSOs follows the distribution of the high column density matter, 
with YSOs clustering at various locations, indicating that recent star formation is going on at multiple sites.

4. We find the possible existence of two generation of massive to intermediate-mass star formation in the complex; 
one is in the immediate vicinity of the Sh2-90 \hii region, in the form of NIR/optical point sources  
responsible for the excitation of compact IR-blobs and a compact \hii region, and the other in the form of young ($\sim$ a few $\times$ 10$^{4}$ yr) Class0/I MYSOs. 

5. From the evolved state of the Sh2-90 \hii region, together with the presence of B-type stars and YSOs embedded  in a thin shell of dense gas close to the IF of the \hii region, we  suggest that the formation of these sources have  possibly been triggered by the expansion of the \hii region. However,  detailed velocity and age measurements of the stars in the \hii region could give more insights into this scenario.

In summary, it appears that multi-generation star formation is going on, but it remains unclear how the star 
formation sites and processes are interlinked.  Taking the present observational evidence at face value, 
we suggest that triggered star formation possibly takes place  at the immediate periphery 
of Sh2-90. However, we hypothesize that the MYSOs currently observed at various locations of the complex 
could have  formed spontaneously or  by some other processes.      
     
\begin{acknowledgements}
We thank the anonymous referee for a critical reading of the paper and
several useful comments and suggestions, which greatly improved the
scientific content of the paper. M. R. Samal acknowledges the financial support provided by the French Space Agency (CNES) for his  postdoctoral fellowship. 
 We thank C. Beaumont and J. Dale for allowing us to use their CO observations and star formation simulation maps, respectively. 
This research has made use of the SIMBAD database operated at the CDS, strasbourg, France, and of the interactive sky atlas Aladin. We acknowledge the support of data analysis facilities provided by the Starlink Project 
which is run by CCLRC on behalf of PPARC. This work used the observations obtained with the {\it Herschel}-PACS and
{\it Herschel}-SPIRE photometers. PACS has been developed by a consortium of institutes led by MPE (Germany) and including UVIE (Austria); KU Leuven, CSL, IMEC (Belgium);
CEA, LAM (France); MPIA (Germany); INAF-IFSI/OAA/OAP/OAT, LENS, SISSA (Italy); IAC (Spain). This development has been supported by the funding agencies BMVIT (Austria), ESA-PRODEX (Belgium), CEA/CNES (France), DLR (Germany), ASI/INAF (Italy), and CICYT/MCYT (Spain). SPIRE has been developed by a consortium of institutes led by
Cardiff Univ. (UK) and including Univ. Lethbridge (Canada); NAOC (China); CEA, LAM (France); IFSI, Univ. Padua (Italy);
IAC (Spain); Stockholm Observatory (Sweden); Imperial College London, RAL, UCL-MSSL, UKATC, Univ. Sussex (UK); Caltech, JPL,
NHSC, Univ. colourado (USA). This development has been supported by national funding agencies: CSA (Canada); NAOC (China); CEA,
CNES, CNRS (France); ASI (Italy); MCINN (Spain); SNSB (Sweden); STFC and UKSA (UK); and NASA (USA).
We thank the French Space Agency (CNES) for financial support. This work is based in part on observations made with the 
{\it Spitzer Space Telescope}, which is operated by the Jet Propulsion Laboratory, California Institute of Technology, under 
contract with NASA. We have made use of the NASA/IPAC Infrared Science Archive to obtain data products from the 2MASS, WISE,
{\it Spitzer}-GLIMPSE, and {\it Spitzer}-MIPSGAL surveys.

\end{acknowledgements}


\newpage 
\begin{table*}
\centering
\scriptsize
\caption{Photometric data of the YSOs in the Sh2-90 complex. A sample of the table is given here. The complete table is available in the electronic form.}
\begin{tabular}{ccccccccccccccccc}
\hline\hline
\multicolumn{1}{c} {RA (deg)} & \multicolumn{1}{c}  {Dec (deg)} &\multicolumn{1}{c}{J} &\multicolumn{1}{c}{eJ} &\multicolumn{1}{c}{H} &\multicolumn{1}{c}{eH} & \multicolumn{1}{c}{K} &\multicolumn{1}{c}{eK}& \multicolumn{1}{c}{[3.6]} &\multicolumn{1}{c}{e[3.6]} &\multicolumn{1}{c}{[4.5]} &\multicolumn{1}{c}{e[4.5]} & \multicolumn{1}{c}{[5.8]} &\multicolumn{1}{c}{e[5.8]}& \multicolumn{1}{c}{[8.0]} &\multicolumn{1}{c}{e[8.0]} & Seq \\

\multicolumn{1}{c} {J2000}  & \multicolumn{1}{c} {J2000} & \multicolumn{1}{c} {mag} & \multicolumn{1}{c} {mag} & \multicolumn{1}{c} {mag} &\multicolumn{1}{c}{mag} &\multicolumn{1}{c}{mag} & \multicolumn{1}{c}{mag}& \multicolumn{1}{c} {mag}& \multicolumn{1}{c} {mag} & \multicolumn{1}{c} {mag} & \multicolumn{1}{c} {mag} &\multicolumn{1}{c}{mag} &\multicolumn{1}{c}{mag} & \multicolumn{1}{c}{mag}& \multicolumn{1}{c} {mag} &  \\
\hline

297.322846 & 26.808891 & 18.726& 0.058 & 16.413& 0.040 & 14.726& 0.018 & 12.809& 0.050 & 11.767& 0.069 & 11.036& 0.073 & 10.341& 0.189 & 1\\
297.300537 & 26.870571 & 17.311& 0.014 & 15.521& 0.014 & 14.382& 0.025 & 12.359& 0.063 & 11.501& 0.071 & 10.848& 0.081 & 9.932& 0.100 & 2\\
297.307587 & 26.838230 & 17.033& 0.090 & 13.603& 0.035 & 11.047& 0.039 & 8.304& 0.041 & 7.304& 0.034 & 6.541& 0.031 & 5.968& 0.021 & 3\\
\hline
\end{tabular}
\end{table*}
 \begin{table*}
\scriptsize
\caption{ Physical parameters of the YSOs derived from the SED fittings}
\begin{tabular}{cccccccccc}
\hline\hline
 ID & \multicolumn{1}{c}{RA(deg)} & \multicolumn{1}{c}{DEC(deg)} &\multicolumn{1}{c}{M$_{\ast}$} &\multicolumn{1}{c}{T$_{\ast}$ } & \multicolumn{1}{c}{  M$_{\rm disk}$ }
& \multicolumn{1}{c}{$\dot{M}_{\rm disk}$ }   &\multicolumn{1}{c}{$L_{\rm bol}$} & \multicolumn{1}{c}{A$_{V}$ }  & \multicolumn{1}{c}{$M_{env}$ } \\

  &\multicolumn{1}{c}{(J2000)} & \multicolumn{1}{c}{(J2000)} & \multicolumn{1}{c}{($M_\odot$)}
          & \multicolumn{1}{c}{(10$^{3}$ K)} & \multicolumn{1}{c}{( $M_\odot$)} & \multicolumn{1}{c}{(10$^{-5}$ $M_\odot$/yr)} & \multicolumn{1}{c}{(10$^{3}$ \lsun)} & \multicolumn{1}{c}{(mag)} & \multicolumn{1}{c}{(\msun)} \\ \hline
 \hline
10 & 297.310202 & 26.836946 & 6.2$\pm$1.1    & 4.2$\pm$0.4     & 0.2$\pm$0.1     & 48.3$\pm$3.2      & 0.7$\pm$0.1     & 16.2$\pm$3.3    & 68$\pm$5.5 \\
13 & 297.304992 & 26.849239 & 7.4$\pm$1.2    & 4.6$\pm$0.3     & 0.5$\pm$0.2    & 33.5$\pm$2.1      & 0.8$\pm$0.1     & 12.4$\pm$1.1    & 19$\pm$2.2 \\
25 & 297.336862 & 26.772658 & 5.8$\pm$1.3    & 3.9$\pm$0.2     & 0.3$\pm$0.2    & 56.0$\pm$3.1      & 0.6$\pm$0.1     & 42.3$\pm$7.8    & 33$\pm$3.6 \\
21 & 297.382867 & 26.753855 & 8.8$\pm$0.8     & 16.4$\pm$4.4    & 0.1$\pm$0.1     & 8.1$\pm$1.1     & 4.0$\pm$0.4     & 31.4$\pm$4.5    & 108$\pm$9.7 \\
\hline
\end{tabular}
\end{table*}
\end{document}